\documentclass[fleqn,usenatbib]{mnras}

\usepackage{newtxtext,newtxmath}

\usepackage[T1]{fontenc}

\DeclareRobustCommand{\VAN}[3]{#2}
\let\VANthebibliography\thebibliography
\def\thebibliography{\DeclareRobustCommand{\VAN}[3]{##3}\VANthebibliography}

\usepackage{graphicx}	
\usepackage{amsmath}	

\newcommand{\Msun}{\text{M}_\odot}
\newcommand{\MpZAMS}{\text{M}_\text{1,ZAMS}}
\newcommand{\MsZAMS}{\text{M}_\text{2,ZAMS}}

\newcommand{\MpiBH}{\text{M}_\text{1,iBH}}
\newcommand{\Msi}{\text{M}_\text{2,i}}

\newcommand{\MpBH}{\text{M}_\text{1,BH}}
\newcommand{\MsBH}{\text{M}_\text{2,BH}}
\newcommand{\qBH}{q_\text{BH}}

\newcommand{\Mdonor}{\text{M}_\text{d}}
\newcommand{\Mac}{\text{M}_\text{a}}

\title[Understanding the high-mass BBH population]{Understanding the high-mass binary black hole population from stable mass transfer and super-Eddington accretion in \textsc{bpass}}

\author[M. M. Briel]{
M. M. Briel,$^{1}$\thanks{E-mail: max.briel@gmail.com}
H. F. Stevance,$^{1,2}$
J. J. Eldridge$^{1}$
\\
$^{1}$Department of Physics, University of Auckland, Private Bag 92019, Auckland, New Zealand\\
$^{2}$Astrophysics Research Centre, School of Mathematics and Physics, Queen’s University Belfast, N. Ireland, BT7 1NN, United Kingdom.\\
}

\date{Accepted XXX. Received YYY; in original form ZZZ}

\pubyear{2022}

\begin{document}
\label{firstpage}
\pagerange{\pageref{firstpage}--\pageref{lastpage}}
\maketitle


\begin{abstract}
With the remarkable success of the LVK consortium in detecting binary black hole mergers, it has become possible to use the population properties to constrain our understanding of the progenitor stars' evolution. The most striking features of the observed primary black hole mass distributions are the extended tail up to $100 \Msun$ and an excess of masses at $35 \Msun$. Currently, isolated binary population synthesis have difficulty explaining these features. Using the well-tested \textsc{bpass} detailed stellar binary evolution models to determine mass transfer stability, accretion rates, and remnant masses, we postulate that stable mass transfer with super-Eddington accretion is responsible for the extended tail. These systems are able to merge within the Hubble time due to more stable mass transfer at higher donor masses with higher mass ratios and spin-orbit coupling allowing the orbits to shrink sufficiently. Furthermore, we find that in \textsc{bpass} the $35 \Msun$ excess is not due to pulsational pair-instability, as previously thought, but a feature caused by stable mass transfer, whose regime is limited by the  mass transfer stability, quasi-homogeneous evolution, and stellar winds. These findings are at odds with those from other population synthesis codes but in agreement with other recent studies using detailed binary evolution models.
\end{abstract}

\begin{keywords}
black hole mergers -- binaries: general -- stars: massive
\end{keywords}



\section{Introduction}

In the years following the first observed binary black hole (BBH) merger in 2015, the number of observations has increased dramatically to a total of 76 BBH mergers with the release of Gravitational Wave Transient Catalogue 3 (GTWC-3; \citealt{abbott_2021c}). This growing population is a window into the stellar physics governing the life and death of their progenitors. However, due to the million to billion year lifetimes, direct observation of their evolution is impossible. Therefore, population synthesis is used to probe the origins of merging BBH systems \citep[see][and references therein]{mandel_2021} with the assumed evolutionary physics leaving imprints on the primary remnant mass, spin, and mass ratio distributions of the merging BBH population. However, GWTC-3 has shown that the BBH population is not yet well understood with the observed primary black hole (BH) mass ($\text{M}_\text{1,BH}$) distribution extending up to $100 \Msun$ - beyond the expected maximum \citep{heger_2002, woosley_2002, woosley_2017, farmer_2019, renzo_2020, woosley_2021, mehta_2022, farag_2022} - and an excess at $35 \Msun$ that is not connected to the turn-off of the distribution. From an isolated binary evolution perspective, the shape of this distribution is mostly influenced by the supernova and mass transfer prescriptions.

Depending on the supernova prescription, the primary remnant mass distribution can be drastically different \citep{mandel_2020, shao_2021, ghodla_2022} with pulsational pair-instability supernovae (PPISN) and pair-instability supernovae (PISN) leaving the largest imprint on the distribution at high black hole masses \citep{spera_2017, marchant_2019, stevenson_2019}. These processes take place when extremely massive stars ($\text{M}_\text{ZAMS} \gtrsim 70 \Msun$; $\text{M}_\text{He} \gtrsim 30 \Msun$) enter a high-temperature, low-density regime in their core allowing for electron-positron pair production. At high core masses ($\text{M}_\text{He} \gtrsim 65 \Msun$), this leads to a single pulsational that completely disrupts the star, the PISN, and leaves no remnant behind \citep{fowler_1964, rakavy_1967, heger_2002}. At smaller core masses, pulsational pair-instability drives mass ejection of the star through multiple pulsations. Eventually, the stellar structure exits the high-temperature, low-density regime and the star continues to evolve normally until it undergoes a core-collapse supernova. The supernova occurring after a pulsational pair-instability phase are referred of which the progenitors have experienced pulsational pair-instability are referred to as PPISN. 
As a consequence of the mass ejection, a large range of Zero-Age Main-Sequence (ZAMS) masses converge to a narrow range of helium core masses between $35 - 50 \Msun$ \citep{woosley_2017}. Depending on the remnant mass prescription, these result in similar BH masses and a pile-up before an absence of them due to PISN \citep{heger_2002, woosley_2002, stevenson_2019, marchant_2019, farmer_2019}. It is, therefore, expected that a pile-up is associated with a cut-off in the primary BH mass distribution. Since more massive stars ($\text{M} \gtrsim 260 \Msun$; $\text{M}_\text{He} \gtrsim 140 \Msun$ ) completely collapse into a BH, the pair-instability disruption creates a gap in the isolated BH mass distribution, also know as the 'upper mass gap' or 'PISN mass gap'. The lower edge of this gap is at $\text{M}_\text{BH} \approx 50 \Msun$ \citep[see][and references therein]{woosley_2021}. This limit can be raised using rapid rotation \citep{marchant_2020, woosley_2021} or altering the nuclear reaction rates \citep{woosley_2021, mehta_2022, farag_2022} to include more massive BHs, which have been observed in the PISN mass gap, such as GW190521. However, the pile-up mass from PPISN increases as well, further decreasing the association with the observed excess at $35 \Msun$.

Without altering the stellar physics, such massive BBH merger could also be explained by non-isolated binary formation pathways \citep{yang_2019, rodriguez_2019, santoliquido_2020, dicarlo_2020, renzo_2020c, mapelli_2021, bouffanais_2021, zevin_2021, arca-sedda_2020a, arca-sedda_2021, costa_2022, ballone_2022, banerjee_2022}. Each would leave imprints on the properties of the merging BBH population and could challenge the standard isolated binary evolution formation pathway.

Another possibility for the existence of BHs in the PISN mass gap is mass transfer onto the BH from a companion \citep{vanson_2020,woosley_2021}. This occurs when the radius of the star, the donor star, expands beyond its Roche Lobe Radius and, thus, mass is transferred to the companion, the accretor. This can drastically alter the mass ratios and evolution of the star with the stability and the accretion efficiency determining the outcome of this mass transfer.

If the radius of the donor star contracts or remains constant as a response to mass loss during the Roche Lobe Overflow (RLOF), the mass transfer is stable. 
If, on the other hand, the star expands as a response to mass transfer, the positive feedback-loop results in the donor star engulfing the whole system in a Common Envelope (CE) phase \citep{paczynski_1976, webbink_1984, iben_1993,podsiadlowski_2001, ivanova_2013}, where part or all of the envelope is removed from the system. The accretion onto the BH during the CE phase is limited \citep{de_2020}, but RLOF before and/or after the CE can result in accretion onto the BH and its effects is poorly explored for high-mass stars.

Systems undergoing CE are, currently, considered the main formation channel for merging BBH systems \citep{dominik_2012, belczynski_2016,bavera_2021b, zevin_2021, zevin_2022, vanson_2022} with only the high-mass primary mass systems being formed through stable mass transfer (SMT) \citep{neijssel_2019, vanson_2022}. However, stable mass transfer (SMT) has been shown to play a more important role than previously thought with accretion being more stable in detailed stellar models than those used in rapid population synthesis \citep{marchant_2021, gallegos-garcia_2021, klencki_2021}, like \textsc{compas} \citep{riley_2022}, \textsc{startrack} \citep{belczynski_2016}, \textsc{mobse} \citep{giacobbo_2018a}, and \textsc{cosmic} \citep{breivik_2020} which use stability criteria based on the evolutionary phase of the star from \citet{hurley_2002}. Using improved stability criteria \cite{olejak_2021a} has shown that more stable mass transfer takes place in \textsc{startrack}. While this results in an extended primary BH mass range up to $55 \Msun$, it is unable to predict more massive primary mass BHs.

Super-Eddington accretion onto the BH during SMT could allow the primary BH to gain a significant amount of mass to become a PISN mass gap BH \citep{vanson_2020}. However, for SMT systems to merge within the Hubble time, they need to lose angular momentum to reduce the size of their orbit. With Eddington luminosity limited accretion, this happens through mass loss from the system during RLOF, but this does not occur for super-Eddington accretion and BHs in the PISN mass gap are unable to merge within the Hubble time \citep{vanson_2020, bavera_2021b, zevin_2022}. Thus, most BBH population synthesis codes limit the accretion onto the BH to the Eddington luminosity. On the other hand, super-Eddington accretion onto a BH is a candidate for Ultra Luminous X-ray sources \citep{woosley_2021} and is theoretically possible \citep{sadowski_2016, woosley_2021}

Since the high end of the primary BH mass distribution is not yet well understood, we predict the properties of the BBH population using detailed stellar models. These models from \textsc{bpass} \citep{eldridge_2017, stanway_2018a} allow for super-Eddington accretion onto a BH and modelling of the response of the donor star to the mass loss without relying on parameterisation for this response. In combination with the star formation history and metallicity evolution from the state-of-the-art TNG-100 simulation, we predict the primary mass and mass ratio distributions of the merging BBH population, and compare them against the observed population. 
In Section \ref{sec:Method}, we discuss \textsc{bpass}, its prescription for mass transfer stability and efficiency, remnant mass, and PPISN prescription. We present our results in Section \ref{sec:results}, where explore the formation pathways of the merging BBH systems and describe the formation of features matching observation in the high BH mass range in Section \ref{sec:primary_mass_features}. In Section \ref{sec:MT_stability}, we discuss how the stability criteria of mass transfer create this outcome. We look at the impact from (P)PISN and quasi-homogeneous evolution on the $\MpBH$ distribution and its features, in Section \ref{sec:robustness}.
In Section \ref{sec:open_problems}, we discuss how this relates to the current view and what limitations our predictions have. We present our conclusions in Section \ref{sec:conclusion}. 


\section{Method} \label{sec:Method}

\subsection{Population synthesis}
To calculate the number of merging BBH systems within the Hubble time, we use the detailed stellar models and population synthesis of Binary Population Synthesis And Spectral Synthesis (\textsc{bpass}) v2.2.1 \citep{eldridge_2017, stanway_2018a}. Here, we will give a short overview of relevant mechanics and changes compared to previous work \citep{eldridge_2019, tang_2020, briel_2022}. We implement improvements made by \citet{ghodla_2022} to rejuvenation and the merger time calculation and use the cosmological star formation history from the TNG100-1 simulation \citep{springel_2018,nelson_2018,pillepich_2018a,naiman_2018,marinacci_2018}, as in \citet{briel_2022}. We use the cosmological parameters from \citet{aghanim_2020} ($h=0.6766$, $\Omega_\text{M} = 0.3111$, and $\Omega_\Lambda = 0.6889$).

\textsc{bpass} is a population synthesis suite that uses a grid of $\sim250,000$ 1D theoretical stellar models that include single and binary evolution and contains 13 metallicities mass fractions ranging from $10^{-5}$ to $0.04$. It uses a custom version of the Cambridge \textsc{stars} code \citep{eggleton_1971}, and is described in \citet{eldridge_2008, eldridge_2017} and \citet{stanway_2018}. It implements metallicity dependent wind mass loss rates from \citet{dejager_1988}, \citet{vink_2001} for OB stars and \citet{nugis_2000} for hydrogen depleted Wolf-Rayet stars.

The fiducial \textsc{bpass} v2.2.1. population is weighted using initial binary parameters from \citet{moe_2017} and a \citet{kroupa_2001} initial mass function extended up to $300 \Msun$. This has been tested against a wealth of observables and can self-consistently reproduce a number of massive star evolutionary characteristics, young and old stellar populations \citep{eldridge_2009, wofford_2016, eldridge_2017, stanway_2018a}, and transients rates \citep{eldridge_2016, eldridge_2019, tang_2020, ghodla_2022, briel_2022}.

\textsc{bpass} implements a unique method in evolving binary systems to reduce computational time. Before the first supernova, the primary star is evolved in detail using the adapted \textsc{STARS} code \citep{eggleton_1971}, while the secondary star is evolved using the single star rapid evolution equations from \citet{hurley_2002}. This is know as the primary model. After the fate of the primary star has been determined and a natal kick has been applied, a secondary model is selected and the secondary star is evolved in detail, either as a single star or as a binary with a compact object, contingent on the outcome of the primary. If the secondary has accreted more than 5 per cent of its initial mass and has a metallicity fraction of 0.004 or below, the secondary model is, instead, replaced by a quasi-homogeneous model (QHE). We explore a more detailed accretion QHE prescription by \citet{ghodla_2022a} in Section \ref{sec:QHE_limit}. Once the secondary also reaches the end of its life, its fate is determined and a natal kick is applied again. If the system is still bound, and contains two compact objects, the coalescence time is calculated using \citet{peters_1964}. We define this as the merger time of the system, while the delay time includes the lifetime of the progenitor stars before the formation of the secondary BH.

\subsection{Remnant mass prescription}
In the fiducial \textsc{bpass} population, the remnant mass is determined by calculating the remaining bound mass after injecting $10^{51}$ erg into the star while taking into account its internal structure \citep{eldridge_2004}. The remnant receives a natal kick from a Maxwell-Boltzmann distribution with $\sigma=265$ km s$^{-1}$ \citep{hobbs_2005} for masses below $3 \Msun$ and is otherwise reduced by a factor $\text{M}_\text{remnant}/1.4\Msun$. Per model we sample 1000 kicks from this distribution to cover the effects of the supernova on the binary.

\begin{figure}
    \centering
    \includegraphics[width=\columnwidth]{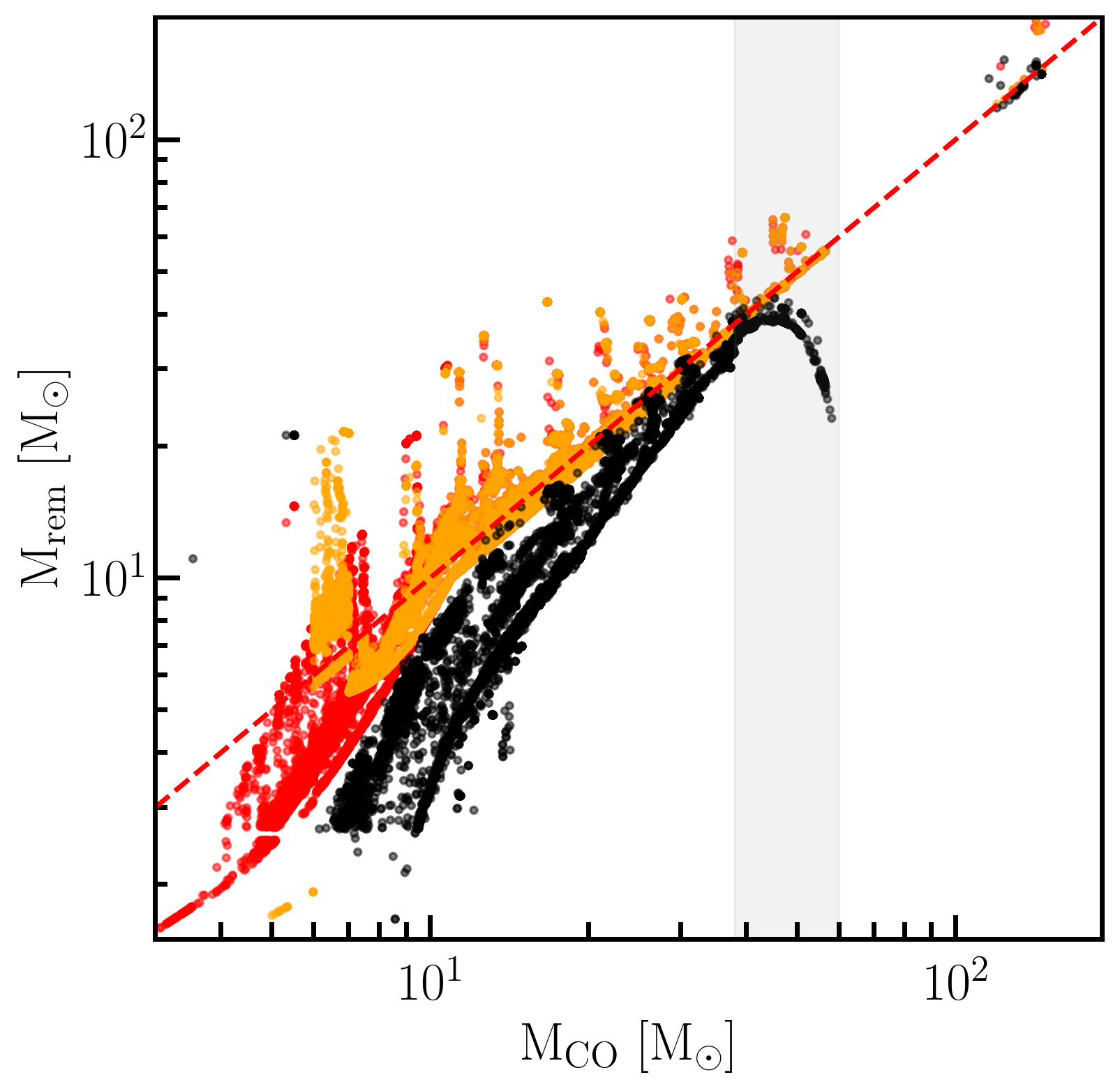}
    \caption{The remnant mass over the CO core for the primary and secondary models at their formation resulting into BBH mergers at $z=0$ using the prescription used in this work (black), and the Rapid (orange) and Delayed (red) prescriptions from \citet{fryer_2012}. At $\text{M}_\text{CO} > 11$, the \citet{fryer_2012} are similar and overlap. The PPISN regime is marked in grey. We indicate where the remnant is the same size as the carbon-oxygen core (red dashed line). Section \ref{subsubsec:fryer} discusses the impact of the \citet{fryer_2012} prescriptions on the $\MpBH$ distribution.}
    \label{fig:CO_core_vs_rem}
\end{figure}

While the standard \textsc{bpass} output does contain a PISN prescription \citet{woosley_2002}, it does not contain a PPISN prescription. This process, however, is essential in determining the BBH population \citep{stevenson_2019, marchant_2019, broekgaarden_2021a}, as its thought to result in the upper mass gap and a pile-up of events. For this work, we implement the prescription from \citet{farmer_2019} based on the CO core of the progenitor, while keeping the standard \textsc{bpass} prescription in the other regimes. Since no upper limit based on the CO core is currently available \citep[for an overview of limits see][]{woosley_2021}, we use the upper limit based on the He core mass from  \citet{woosley_2002}. This results in the prescription in Equation \ref{eq:remnant_prescription}, which transitions smoothly between regimes, as shown in Figure \ref{fig:CO_core_vs_rem}.

\begin{align*}
    \text{M}_\text{CO} < 38\ \Msun  \longrightarrow &\: 10^{51} \text{erg injection} \\
    38\ \Msun \leq \text{M}_\text{CO} < 60\ \Msun \longrightarrow &\: \text{\citet{farmer_2019}} \\
    \text{M}_\text{CO} \geq 60\ \Msun \text{ and } \text{M}_\text{He} < 133\ \Msun \longrightarrow &\: \text{M}_\text{rem} = 0 \\
    \text{M}_\text{He} \geq 133\ \Msun \longrightarrow &\:
    10^{51} \text{erg injection}
    \addtocounter{equation}{1}\tag{\theequation}
    \label{eq:remnant_prescription}
\end{align*}

Figure \ref{fig:CO_core_vs_rem} shows the relation between CO core and remnant masses for the primary and secondary models resulting in BBH mergers. In Section \ref{subsec:remnantmass}, we explore the effect of the \citet{fryer_2012} prescriptions and (P)PISN on the $\MpBH$ distributions. Furthermore, we change baryonic masses into gravitational masses, following \citep{fryer_2012} with $\text{M}_\text{grav} = 0.9 \text{M}_\text{bar}$ for remnant masses above $3\Msun$ and for masses below we use  

\begin{equation}
    \text{M}_\text{grav} = 2 \frac{\sqrt{1 + 3\text{M}_\text{bar} -1}}{3}
\end{equation}

\subsection{Mass transfer}
\label{sec:MT}
Mass transfer is one of the most important effects in binary evolution with about 50-70 per cent of massive stars experiencing it during their lifetime \citep{sana_2012, sana_2013}. Determining the stability of this mass transfer is challenging, because it requires detailed modelling of the structure of the star to determine its response to the mass loss $\zeta = \text{d}\log R/\text{d}\log M$. To this end, rapid population synthesis codes use approximations for this response based on the evolutionary phase of the star \citep{belczynski_2020,olejak_2021, riley_2022}. However, \textsc{bpass} uses detailed stellar models that calculate the equations describing the stellar interior in a time resolved manner, thus we directly model the response of the donor radius, which leads to significant differences in the donor response \citep{woods_2011, pavlovskii_2015}.
When a star fills its Roche Lobe, mass is transferred based on the amount of overflow according to \citet{hurley_2002}: 
\begin{align}
\label{eq:MT}
    &\dot{\text{M}}_\text{d} =  F(\Mdonor)[\ln(\text{R}_\text{d}/\text{R}_\text{RL})]^3 \Msun \text{yr}^{-1} \\
    \text{with } F(\Mdonor) &=  3 \times 10^{-6}[\text{min}(\Mdonor, 5.0)]^2
\end{align}

Equation \ref{eq:MT} shows the mass loss rate from the donor star with its mass as $\Mdonor$, and its radius and Roche Lobe radius as $\text{R}_\text{d}$ and $\text{R}_\text{RL}$, respectively. The material in this process can be accreted by a companion star, but it is limited to $\dot{\text{M}}_\text{a} \leq \Mac/\tau_{KH}$, where $\Mac$ is the mass of the accretor star. A compact object below $3 \Msun$ \citep{chamel_2013} is limited to the Eddington luminosity, as per \citep{cameron_1967}. The accretion is unrestricted for masses above this limit, since the accretion onto a BH is uncertain and  multiple methods for super-Eddington accretion have been proposed \citet[see][and references therein]{wiktorowicz_2015}. Additional material is lost to the system with its angular momentum by treating this as a wind from the donor star.

When RLOF takes place, we also consider tidal forces due to the interaction. Due to tidal locking, they can spin up the components of the binary using angular momentum of the orbit, which can result in significant orbital shrinkage depending on the properties of the binary. In some cases, a runaway effect causes the system to merge, also known as the Darwin Instability \citep{darwin_1879}.

If, due to the response of the donor star to mass loss or due to the tidal interactions, the donor star radius expands past the orbital separation of the system, we assume a CE takes place (for a more detailed description of our CE prescription, see Stevance et al. in prep.). In our CE prescription, we remove mass from the star as quickly as the detailed model allows, and relate the binding energy of the material lost to the orbital energy. This process conserves angular momentum, takes into account the structure of the stars, and does not require the envelope to be completely stripped. However, it results in longer duration CE phases, albeit still significantly shorter than the thermal timescale of the stars and should not influence the predictions of the binary evolution \citep{eldridge_2017}. Once the radius shrinks below the orbital separation the CE phase ends. Our CE treatment is close to the $\gamma$ formalism \citep{nelemans_2000, nelemans_2005} due to the angular momentum conservation, as reported in Stevance et al. (in prep.), and we retroactively calculate the $\alpha_\text{CE} \lambda$ from our detailed models for comparison against the $\alpha\text{-}\lambda$ formalism used in rapid population synthesis codes \citep[][]{ivanova_2013}. This results in equivalent values for $\alpha_\text{CE} \lambda$ ranging from 3 to 30 (\citealt{eldridge_2017}; Stevance et al. in prep.). Since $\alpha_\text{CE}$ is a free parameter based on the conversion efficiency of binding energy to orbital energy and $\lambda$ a parameter to determine how the structure of the star affects the binding energy, we do not separate them. Before and after the CE phase, mass can be transferred onto the companion, but accretion is non-existent during the CE. Mass transfer continues until its radius shrinks to below the Roche Lobe radius.


\section{Formation Pathways} \label{sec:results}

\subsection{Cosmic Merger Rate over Redshift}
\begin{figure}
    \centering
    \includegraphics[width=\columnwidth]{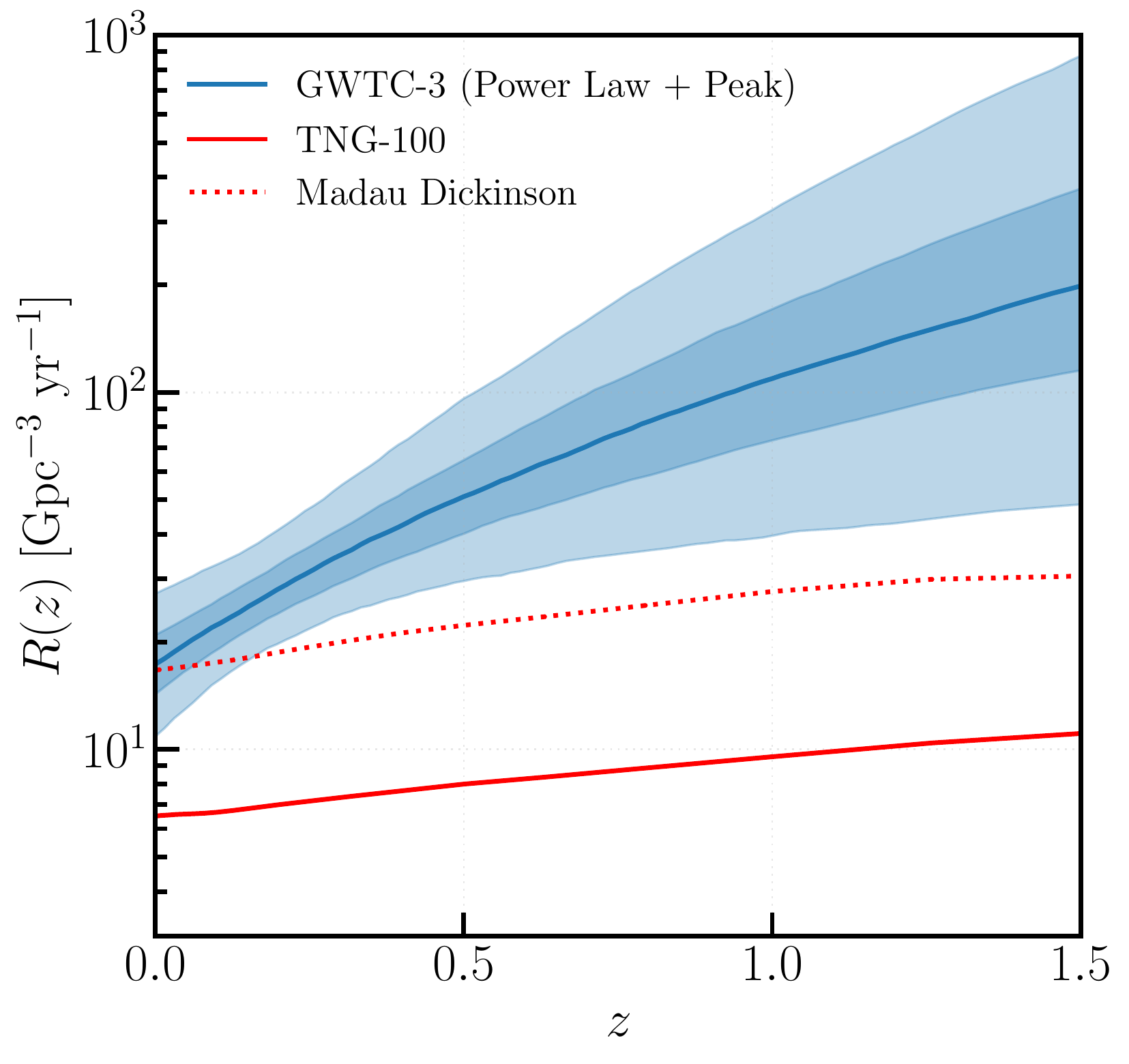}
    \caption{The predicted evolution of the intrinsic BBH merger rate (red solid) over redshift using the TNG-100 cosmic star formation history compared to found evolution found by \citet{abbott_2021c} using their over redshift evolving \textsc{PP} model (blue) with 50 per cent (dark blue) and 90 per cent (light blue) credible bounds. The BBH merger rate using the \citet{madau_2014} cosmic star formation history (red dashed) with the same BBH population as TNG100 is shown to indicate the error from the star formation history.}
    \label{fig:cosmic_event_rate}
\end{figure}

From the intrinsic properties of the merging BBH population of the GWTC-3 analysis, \textsc{bpass} can predict the total cosmic rate of BBH over redshift, the mass ratio distribution and primary mass distribution. In Figure \ref{fig:cosmic_event_rate}, we show the redshift evolution of the BBH merger rate over redshift against the observed intrinsic rate from their \textsc{power law + peak} (\textsc{PP}) population model. It consists of an over-redshift-evolving merger rate power-law and a Gaussian. The predicted rate using the TNG-100 star formation history results in a lower rate than observations. However, it is important to keep in mind that the rate is not only dependent on the assumed physics, but also the cosmic star formation history, which can have drastic effects on the final calculated cosmic rate \citep{chruslinska_2019a, neijssel_2019, santoliquido_2021,broekgaarden_2021a}. We use the same population data, but use the empirical star formation history prescription from \citet{briel_2022}, which uses the \citet{madau_2014} star formation history and the \citet{langer_2006} metallicity evolution, and find that the rate at $z=0$ aligns with observations. However, the evolution over redshift remains shallower than found by \citet{abbott_2021c}.

It is promising that the TNG-100 rate is below the observed rate, since our prediction only includes BBH mergers from isolated binary evolution. Other processes, such as triples and dynamic mergers, are likely to contribute to the observed rate, which are absent in our prediction. These could add at least 30 per cent to the total BBH observed rate \citep{zevin_2021}. However, care should be taken with formation channel predictions due to uncertainties in star formation and stellar physics \citep{broekgaarden_2021a}. 
 
The current TNG-100 BBH merger rate is lower than reported in \citet{briel_2022}, which is due to the improved merger time calculation and the treatment of rejuvenation during mass transfer, as described in \citet{ghodla_2022}. The latter provides a smoothing of the BBH delay time distribution, while the former generates more accurate merger time predictions. Together, these improvements increase the merger time for some systems and, thus, reduce the BBH mergers within the Hubble time. We report the BBH, black-hole neutron-star, and binary neutron star merger rates at $z=0$ in Table \ref{tab:rates} for completion. Except for the BBH rate, these align with current observational constraints from \citet{abbott_2021c}.

\begin{table}
\centering
\begin{tabular}{c|ccc}
      & \textbf{BBH} & \textbf{BHNS} & \textbf{BNS}\\
    & \multicolumn{3}{|c|}{[yr$^{-1}$ Gpc$^{-3}$]} \\
\hline
Predicted   & 6.5 & 27.0 & 93.1 \\
\citet{abbott_2021c} & 10 -- 130 & 7.4 -- 320 &  13 -- 1900\\
\end{tabular}
\caption{The compact object rates at $z=0$ for the TNG-100 star formation history. The observed rates from \citet{abbott_2021c} are the combined 90 per cent confidence intervals for their models for the non-evolving and redshift evolving merger rates.}
\label{tab:rates}
\end{table}

\subsection{Population Properties} \label{subsec:population}
\begin{figure*}
    \centering
    \includegraphics[width=\textwidth]{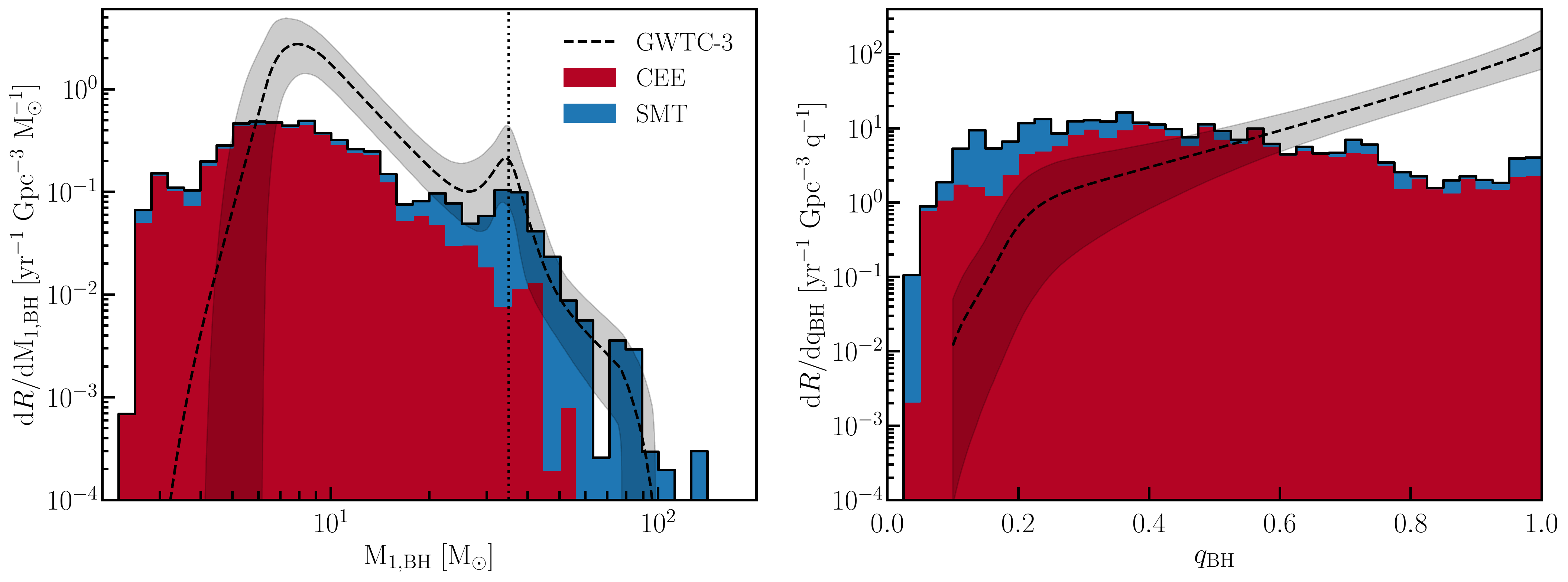}
    \caption{The event rate density over the primary BH mass (left) and the mass ratio (right) at $z=0$ with the results from the \textsc{PP} model from \citet{abbott_2021c} as the dashed line. The gray area is the 95 per cent confidence interval. The presence of a common envelope in the evolution of the BBH merger progenitor has been marked with red, while blue indicated that no common envelope has taken place and only stable mass transfer has occurred during the life of the BBH progenitor system. The vertical dotted line indicates the $35 \Msun$ BH mass. $\mathrm{q}_\mathrm{BH}$ is split into 40 linear bins between 0 and 1.}
    \label{fig:m1_q_distributions}
\end{figure*}

More understanding can be gained from the properties of the population. While observations are still limited, one has to be cautious drawing strong conclusions from these comparison, but the most constraint observables from GWTC-3 are the primary mass and mass ratio distribution. In Figure \ref{fig:m1_q_distributions} we compare the predicted population against the intrinsic observed population from the \textsc{pp} model from \citet{abbott_2021c}. We split the $\MpBH$ space into 40 logarithmic spaced bins between 2 and 200 $\Msun$. Unless otherwise stated we use this binning for all $\MpBH$ distributions. In Appendix \ref{app:uncertainties}, we perform a Poisson error calculation, bootstrap sampling, and apply a kernel density function to determine the uncertainty associated with this distribution, which is limited.

While the predicted rate over mass ratio with $\qBH =\MsBH/\MpBH$, where $\MsBH$ is the mass of the secondary BH in Figure \ref{fig:m1_q_distributions} does increase rapidly at low $\qBH$, it does not continue to increase with increasing $\qBH$ and instead drops beyond $\qBH \sim 0.4$. This might be due to our CE prescription, which dominates the mass ratio distribution, or the remnant mass prescription (for the latter, see Section \ref{subsec:remnantmass}).

Although the mass ratio distribution does not agree with the observed values, we find good agreement in the high-mass regime of the $\MpBH$ distribution between our predictions and GWTC-3. The predicted distribution contains a double peak structure with a peak near $5 \Msun$ and $35 \Msun$ with the latter agreeing with observations. Moreover, it extends into the PISN mass gap with masses over $100 \Msun$, as shown in Figure \ref{fig:m1_q_distributions}. However, it does over predict the low mass BBH mergers ($\MpBH \lesssim 5 \Msun$) and underestimates the number of systems near $10\Msun$.

The well-matched nature of the predictions to observations in the high BH mass regime raises questions about the formation of the high-mass BHs and their merger time. Moreover, the peak at $35 \Msun$ seems to be disconnected from the turn-off of the $\MpBH$ distribution, as observed; this calls into question its connection to PPISN, which is thought to lead to a pile-up at the end of the primary BH mass distribution near $ \sim 40-50 \Msun$ \citep{marchant_2019, renzo_2020b, woosley_2021}. Because these questions are linked, the Section \ref{sec:primary_mass_features} will cover these components in detail, but here we will discuss the formation channels giving rise to the $\MpBH$ distribution.

To this end, we tag BBH merger progenitor model that undergo CE during their evolution as CEE, and models that only experience SMT as SMT. If multiple mass transfer phases take place during the evolution, we check if any of them is unstable, if so we tag the model as undergoing CEE. If no interactions take place, the model is tagged as NON. We perform the tagging for the primary and secondary models.

In Figure \ref{fig:m1_q_distributions}, we separate the $\MpBH$ distribution into a formation channel with a CE in their evolution, and in a pure SMT sample, that only contains systems (primary and secondary models) with stable mass transfer. We find that high-mass BH ($20 - 115 \Msun$) are increasingly formed through systems only undergoing SMT, which is in agreement with \citet{neijssel_2019} and \citet{vanson_2022}. The majority of these events have mass ratios below $\qBH <0.4$.

\begin{figure*}
    \centering
    \includegraphics[width=\textwidth]{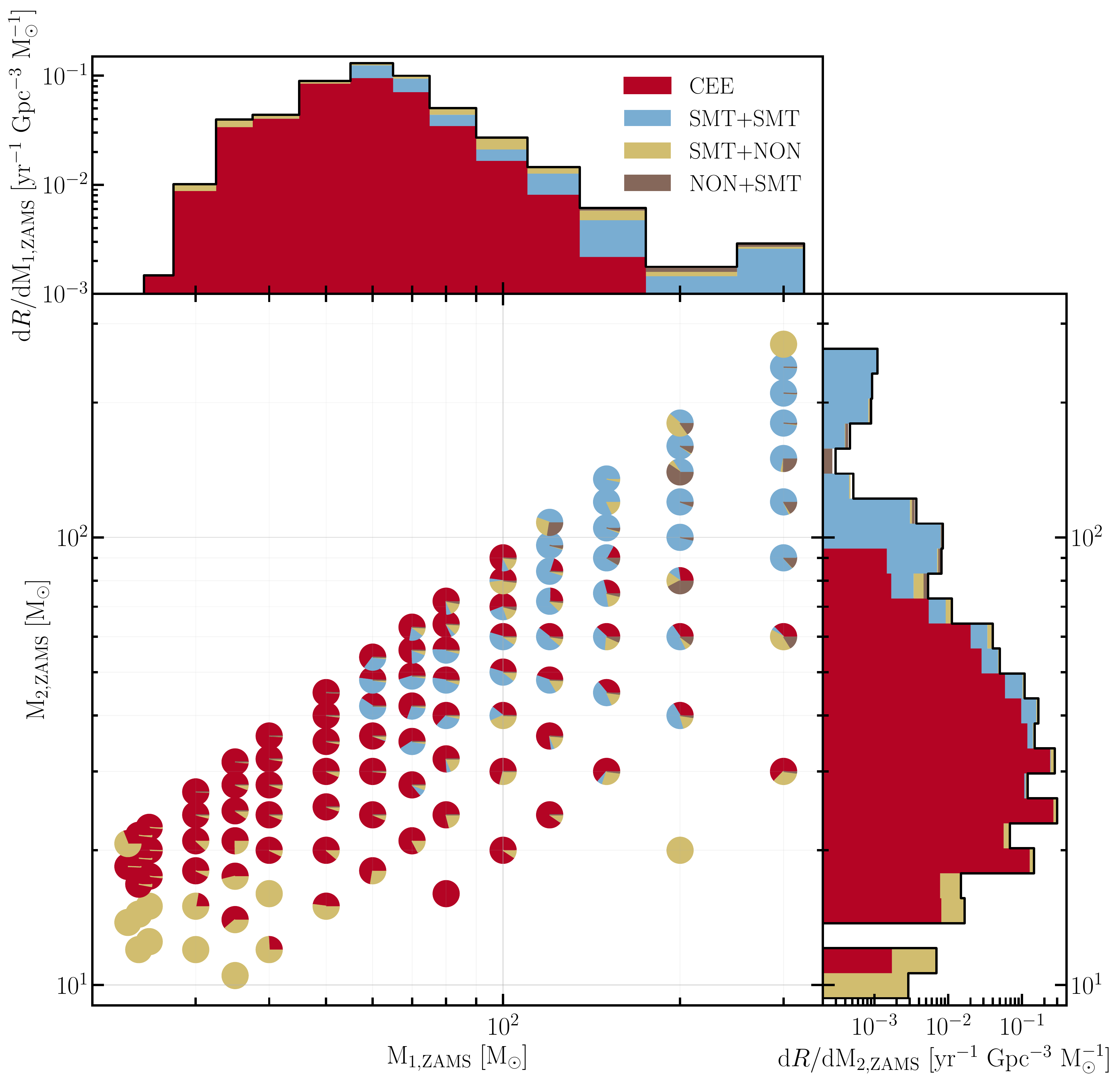}
    \caption{The ZAMS distributions of $M_1$ and $M_2$ for systems leading to a BBH merger at $z=0$. The SMT channel has been split into its individual formation channels, since it dominates the high $\MpBH$ regime. Each circle contains the systems at those $\MpZAMS$ and $\MsZAMS$ and are split into their formation pathways, weighted according to the systems weight w.r.t. total weight at this $\MpZAMS$ and $\MsZAMS$. $\MpZAMS$ is split into 16 bins, which are the ZAMS masses in the \textsc{bpass} grid. $\MsZAMS$ is split into 40 bins between 2 and 300 $\Msun$.}
    \label{fig:ZAMS_SMT}
\end{figure*}

To see what models these systems come from, Figure \ref{fig:ZAMS_SMT} shows that the main channels contributing to the SMT only channel are SMT+SMT with SMT in the primary and secondary model, and SMT+NON, where after SMT in the primary model no interaction in the secondary model takes place. Both channels originate from massive stars ($\MpZAMS > 60 \Msun$) with a slightly less massive companion ($\MsZAMS > 35 \Msun$). This is further motivated by the central figure, which shows the fraction of events going through a formation pathway at a given $\MpZAMS$ and $\MsZAMS$. While these systems originate from similar populations, they undergo drastically different evolution. The NON+SMT systems only contribute  0.30 per cent to the total merging BBH rate, but originate from a wide variety of initial masses.

\begin{figure*}
    \centering
    \includegraphics[width=\textwidth]{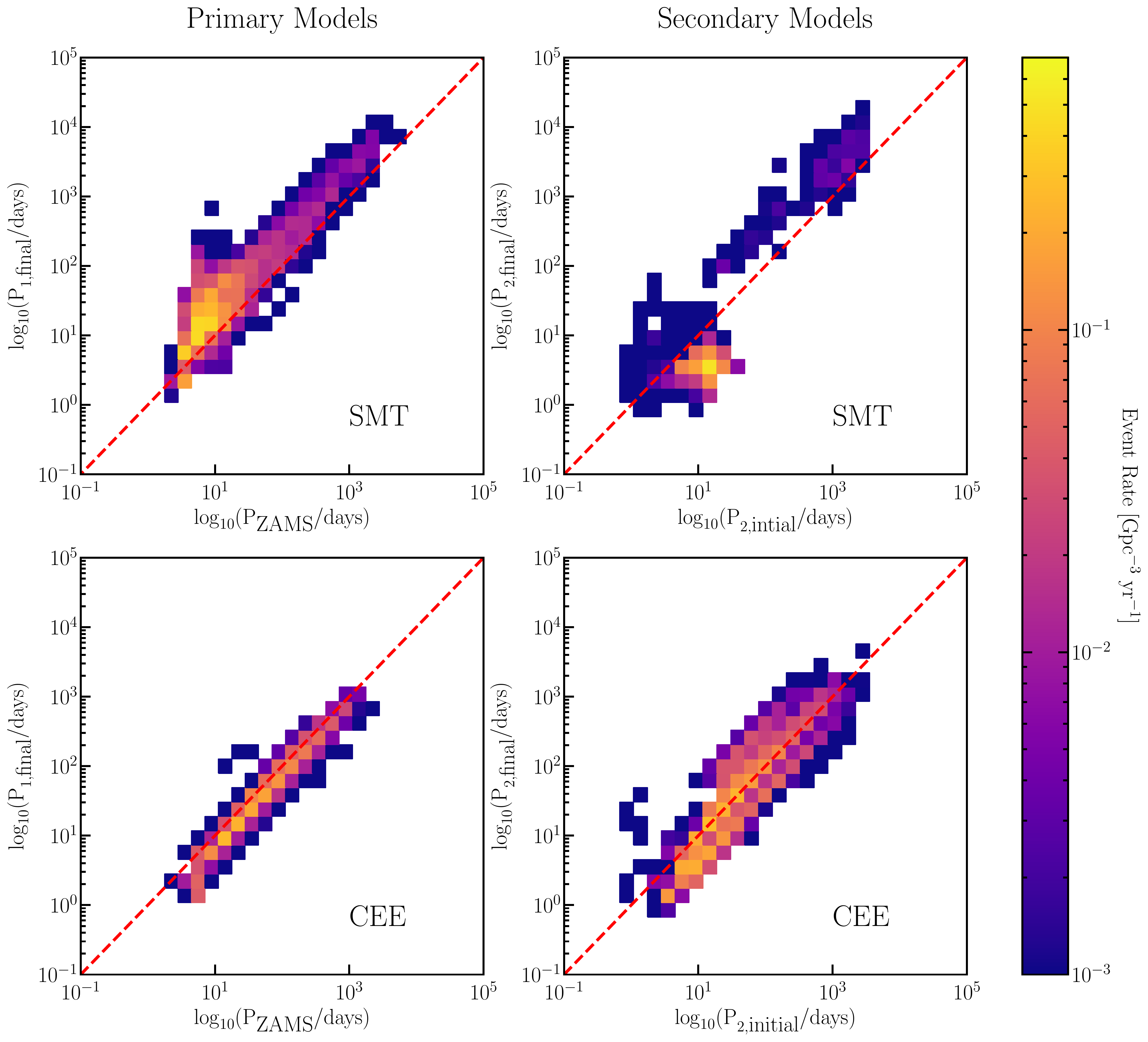}
    \caption{The period evolution of the merging BBH progenitors. The left column contains the primary binary models in the simulation, i.e. the evolution before the primary supernova. The right column contains the secondary binary models after the first supernova. The x-axis represents the period at the start of the model. This means that the P in the left column represents the ZAMS period of the systems. They y-axis represents the period at the end of the model evolution, which does not include the SN. Between the primary (left) and secondary (left) binary models the remnant mass prescription is applied to the primary star. The primary ($1$) and secondary ($2$) stars defined at the beginning of the evolution and we do not change this definition throughout the evolution. The final mass ratio is not the BBH mass ratio, since the remnant mass prescription still needs to be applied to the secondary star.
    The rows are split between SMT and CEE taking place in this model. The periods are logarithmic binned in 30 bins between $10^{-1}$ and $10^{5}$., The dashed red line indicates that the period has remained unchanged throughout the evolution, e.g. $\text{P}_\text{initial} = \text{P}_\text{final}$.}
    \label{fig:P_SMTvsCEE}
\end{figure*}

Nearly all BBH merger progenitor systems interacted during their lifetime and the CE and SMT channel result in different remnant mass outcomes. Figure \ref{fig:P_SMTvsCEE} shows the period evolution separated per evolutionary phase and mass transfer case. In the left column, i.e. the evolution of the binary before the first supernova where the primary star, $\text{M}_1$, initiates mass transfer, while the right column shows the period evolution of the secondary binary models, where the initial primary star has become a compact object and the secondary star, $\text{M}_2$, fills its Roche Lobe. Throughout the evolution we keep the definition of the primary (1) and secondary (2) the same, as such in the primary (secondary) models $\text{M}_1$ is $\Mdonor$ ($\Mac$) and $\text{M}_2$ is $\Mac$ ($\Mdonor$).

\begin{figure*}
    \centering
    \includegraphics[width=\textwidth]{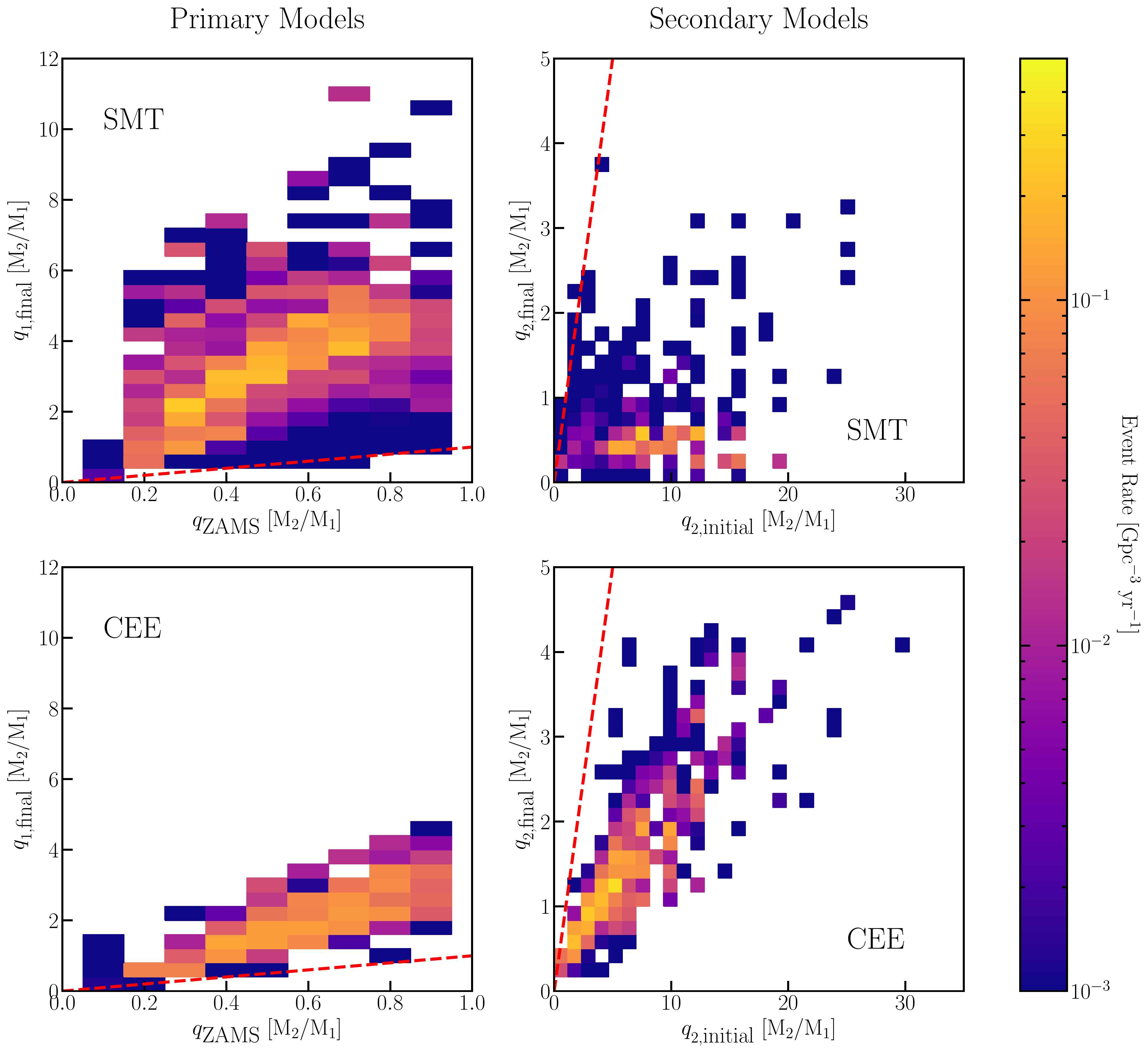}
    \caption{Same as Figure \ref{fig:P_SMTvsCEE}, but for the mass ratio. Depending on the $q$ value the binning is different. The $q_\text{ZAMS}$ bins are centred around the $q$ values used by \textsc{bpass}, 0.1 to 0.9 with steps of 0.1. $q_\text{1,final}$, $q_\text{2,initial}$, and  $q_\text{2,final}$ each contains 30 bins linearly spaced between $0\text{--}12$, $0\text{--}35$, and $0\text{--}5$, respectively. The dashed red line indicates that the period has remained unchanged throughout the evolution, e.g. $q_\text{initial} = q_\text{final}$.}
    \label{fig:q_SMTvsCEE}
\end{figure*}

In the initial phase of the evolution (left), we see that SMT (top) leads to larger periods and, thus larger separations than systems experiencing CEE (bottom). This is a result of material being transferred from the more initially massive star to the less massive in the system. At the start of mass transfer, this results in an orbit shrinkage, but once the mass ratio becomes more equal the orbit starts to widen \citep{soberman_1997, vanson_2020}.

Looking at the mass ratio evolution of SMT systems during the primary phase of the binary evolution in Figure \ref{fig:q_SMTvsCEE} (top left), we see that the mass ratios flip and increase up to $q_\text{1,final}=\text{M}_2/\text{M}_1 = 11$ when the primary star dies,  with most systems laying between 2 and 6. The CEE channel (bottom left) is only able to achieve a maximum mass ratio of $q_\text{1, final} = 5$ with most models between 0.5 and 4, but does reduce the periods for most systems, as expected.

The remnant mass prescription further increases the mass ratio up to $q_\text{2, i}=30$ by reducing the primary mass, as can be seen in the right column in Figure \ref{fig:q_SMTvsCEE}. This column contains the secondary models split between those experiencing SMT and CE, thus, a model in SMT could have experience SMT, CE, or not interacted in the primary model. Most of the systems undergoing only SMT after the first supernova have an initial mass ratios between $5 < q_\text{2,i} < 20$. A few low contributing systems have more extreme mass ratios up to $q_\text{2, i}=25$, which are unexpected in the context of SMT and we will explore these further in Section \ref{sec:q_explosation}. Eventually, the mass transfer stops and the systems reach mass ratios between 0 and 1, which will be further reduced by the remnant mass prescription. Thus, super-Eddington accretion on to the BH leads to a mass reversal, such that the $\MpZAMS$ also becomes $\MpBH$. This results in limited mass ratio reversal \citep{zevin_2022, broekgaarden_2022, mould_2022}, as will be discussed in Section \ref{subsec:MRR}.

The common envelope in the second phase of the systems evolution decreases the mass ratio linearly, e.g. a higher initial mass ratio leads to a higher final mass ratio. Since we tag a system as CEE if the undergoes a single CE phase during the phase of the systems evolution, we do not identify multiple mass transfer phases in the system, which occur in our model. Therefore, phases of SMT in a system undergoing CE can alter the mass ratio and periods.

In general, systems experiencing only SMT in the secondary model, have on average smaller periods at the end of the model than those that have experienced CE. The reason for the orbital shrinkage depends on the binary model parameters, but most shrinkage is caused by tidal synchronisation reducing the angular momentum in the orbit. Most systems are below the critical mass ratio for the Darwin Instability ($q > 12$) \citep{eggleton_2011}, but those above are able to avoid it due to rapid changes in the mass ratio \citep{stepien_2011}, the rapid shrinkage of the donor radius, and/or increased winds due to high helium surface abundance. We further discuss the stability of these models in Section \ref{sec:MT_stability}.

The right column of Figure \ref{fig:P_SMTvsCEE} shows that the period change in during the CEE interaction spreads out its original distribution and reduces the period for systems with period $\lesssim 10^{1.5}$. 
The SMT channel reduces the period of such systems more significantly than the CEE channel. Furthermore, two main clusters of period can be identified around $10^1$ and around $10^3$ with limited systems in between these period. This is a result of interaction on the Hertzsprung gap being mostly unstable and will be discussed more in Section \ref{sec:MT_stability}.
However, both channels have systems with final periods that are unable to merge when assuming a circular orbit, e.g. $10^4$ days. We find that these systems come from higher metallicity populations, and have eccentricities near unity due to the natal kick they received, as such they are able to merge within the Hubble time. This shows that it is possible to get BBH systems merging within the Hubble time, even when considering super-Eddington accretion, while retaining a significant amount of mass, thus resulting in more massive BBH systems.

\begin{figure*}
    \centering
    \includegraphics[width=\textwidth]{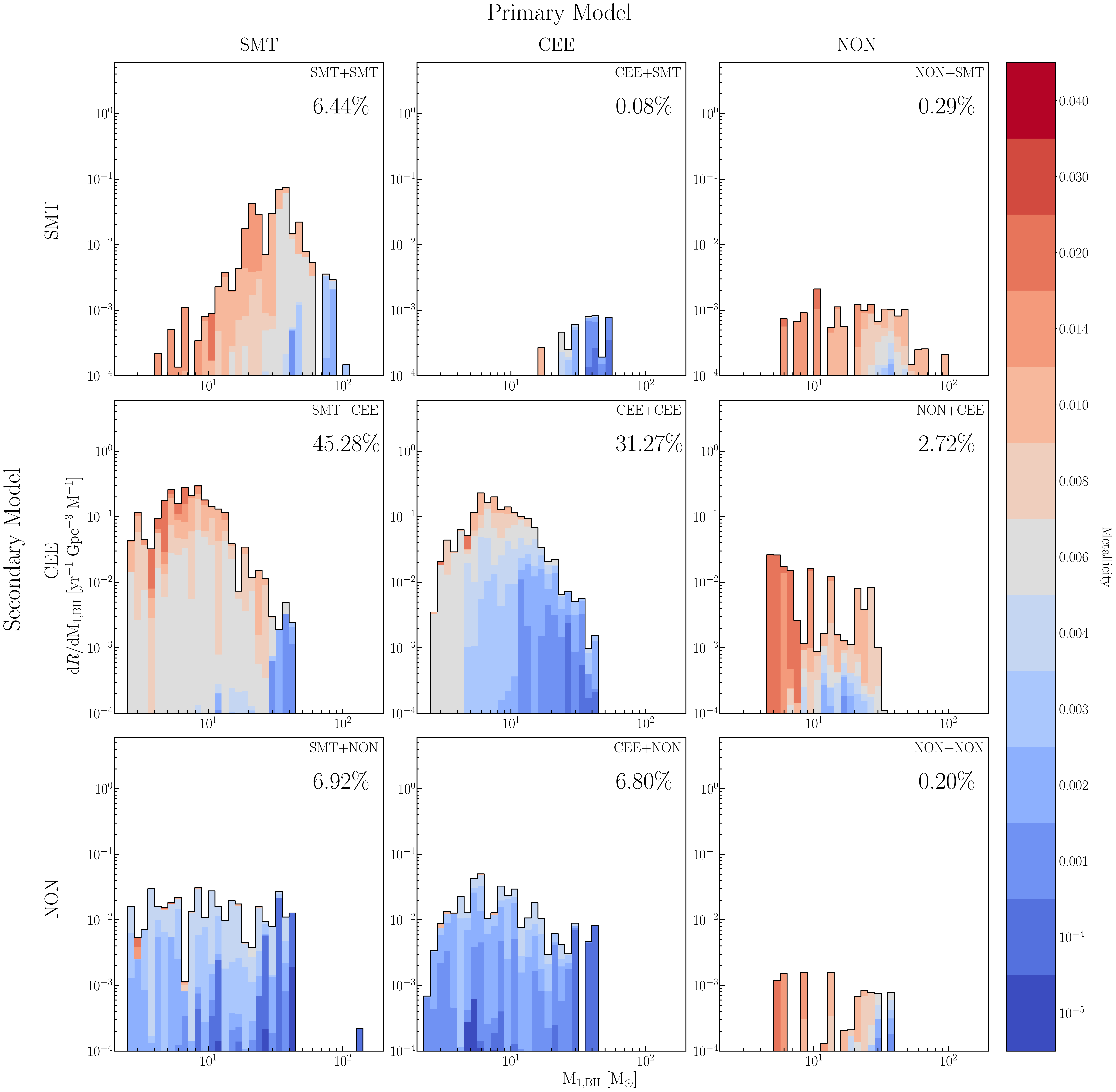}
    \caption{The predicted event rate at $z=0$ over primary BH mass separated by formation channels coloured by the metallicity at which the binary system was formed. The column are split based on only SMT, CEE, or NON occurring in the primary model, as described in Section \ref{subsec:population}. The rows are the same, but for the secondary models. The percentages is the contribution of that formation channel to the total BBH merger rate.}
    \label{fig:formation_channels_met}
\end{figure*}

We further separate the formation pathways per metallicity and per evolutionary model in Figure \ref{fig:formation_channels_met}. The highest contribution comes from the \textsc{SMT+CEE} (45.24 per cent), where the primary model only experiences SMT and the secondary contains a CE phase. The next largest formation channel is \textsc{CEE+CEE} (31.17 per cent) channels, where a CE takes place in both the primary and secondary model. The dominance of CE in the formation of merging BBHs is in agreement with other population synthesis codes \citep{dominik_2012, belczynski_2016, bavera_2021b, zevin_2021, zevin_2022, vanson_2022}. The \textsc{SMT+CEE} distribution contains mergers originating from a variety of metallicities, but lower mass primary mass systems originate from higher metallicities than the higher primary mass systems due to the mass loss from stellar winds. The \textsc{CEE+CEE} distribution, on the other hand, has a higher contribution from low metallicity star formation. 
Following these high contributing events is the \textsc{SMT+NON} channel, where the binary only interacts before the first supernova (6.95 per cent). This distribution is dominated by mergers from low metallicity ($Z < 0.006$) star formation with a large contribution between 30 to 40 $\Msun$ from $Z=0.0001$. While this channel contains the largest primary BH mass, most other events are constrained to $\text{M}_1 < 40 \Msun$. The next largest formation pathway, the \textsc{CEE+NON} channel (6.79 per cent), is also limited to $\text{M}_1 < 40 \Msun$ and low metallicity. Similarly, \textsc{CEE+SMT} only occurs at low metallicity, but has a specific mass regime (15 to 55 $\Msun$) and only contributes 0.08 per cent. The events only interacting in the primary preferentially come from low metallicity events, while systems interacting only in the secondary phase originate from higher metallicity environments. The \textsc{SMT+SMT} channel (6.55 per cent) forms most of the high primary mass BH systems, similar to the distribution shown by \citet{vanson_2022}. We note a metallicity dependence in the primary mass with low primary mass BHs originating from higher metallicity environments, as a result of the formation pathway discussed in Section \ref{subsec:excess}.

\section{Primary remnant mass distribution features} \label{sec:primary_mass_features}
Mass transfer is a defining feature for the outcome of the binary system and stable mass transfer leads to higher mass systems. We find a distribution that matches the high-mass properties of the observed intrinsic primary mass distribution. These are the extended mass range and an excess at $35 \Msun$. In the next sections, we cover these two findings in more details. 

\subsection{The $35 \Msun$ Excess} \label{subsec:excess}
In Figure \ref{fig:m1_q_distributions}, we find an excess of systems near M$_1 \approx 35 \Msun$, similar to observations, which previously has been attributed to a pile-up of PPISN events. We tag the systems undergoing PPISN and find that only the high end of the peak is a result of the PPISN.

\begin{figure}
    \centering
    \includegraphics[width=\columnwidth]{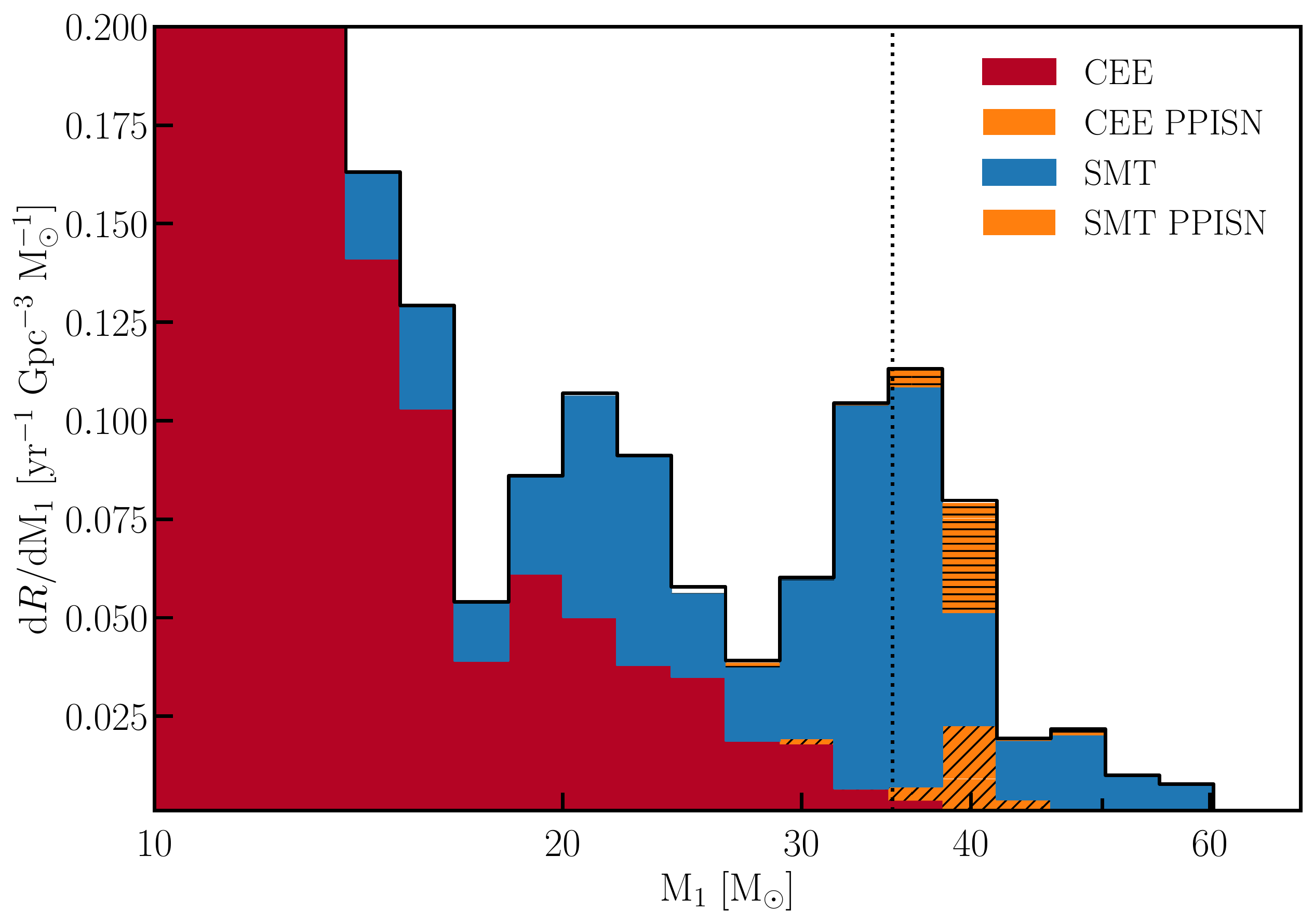}
    \caption{The primary remnant mass distribution at $z=0$ rebinned to 50 logarithmic spaced bins and zoomed in on the $35\Msun$ excess, where the systems are split into the SMT and CEE channels with the PPISN tagged per channel. The systems experiencing PPISN are above $35 \Msun$ (dashed line). The majority of the peak if formed through the SMT channel. The black line represent the total event rate, which also contains non-interacting systems. An additional excess near $20 \Msun$ can be identified and is discussed in Section \ref{subsec:additional_structure}.}
    \label{fig:PPISN_formation_channels}
\end{figure}

As shown in Figure \ref{fig:PPISN_formation_channels}, the majority of $\sim 35\Msun$ BH progenitors do not experience PPISN. Instead, we find that the peak consists of systems only undergoing SMT with PPISN being present around $40 \Msun$ bin. The CEE channel, on the other hand, has a decreasing rate until $35 \Msun$, after which it slightly increases due to PPISN events. Since other population synthesis codes are dominated by CEE systems, the pile-up is logically attributed to PPISN. However, since in this mass regime \textsc{bpass} is dominated by SMT, this is not the main formation process for the $35 \Msun$ peak.

The main contributors to the $35 \Msun$ peak within the SMT channel are the SMT+SMT and SMT+NON formation pathways. These channels originate from massive stars ($M_1 > 60 \Msun$) with a slightly less massive companion ($M_2 > 35 \Msun$), as shown in Figure \ref{fig:ZAMS_SMT}. While these systems originate from similar populations, they undergo drastically different evolution. Therefore, we look at each pathway separately in the following sections.

\subsubsection{Single stable mass transfer only (SMT+NON)}

Because these systems only interact before the first supernova, the first formed BH remains unchanged in mass after its formation and, thus, come from initially massive stars that have experienced or are close to the lower limit of PPISN.

The non-interacting nature after the first SN in this system makes sure that the primary BHs do not continue to grow after their formation. This is a result of the companion experiencing QHE which occurs in \textsc{bpass} when 5 per cent of its initial mass being accreted from the primary at low metallicities. This explains the low average metallicity of this channel in Figure \ref{fig:formation_channels_met}, since QHE is limited to systems with a metallicity of 0.004 or below. The majority of systems in this channel undergo QHE. Our QHE models are assumed to be fully mixed during hydrogen burning, thus their mean molecular weight increases with the progression of nuclear burning. This leads to the radius of the star shrinking rather than expanding, as would occur with normal main sequence stars. Thus, QHE stars fail to fill their Roche Lobe and do not interact with their companion. The BH companion will be unable to grow further and PPISN systems only contribute to the high end of the peak near $35\Msun$ through the SMT+NON formation channel.


\subsubsection{Double stable mass transfer only (SMT+SMT)}

The SMT+SMT formation pathway is more complicated than the SMT+NON channel due to interactions in multiple evolutionary phases. Similar to SMT+NON these merging BBHs come from very massive ZAMS stars, but they do not all experience PPISN.

\begin{figure*}
    \centering
    \includegraphics[width=\textwidth]{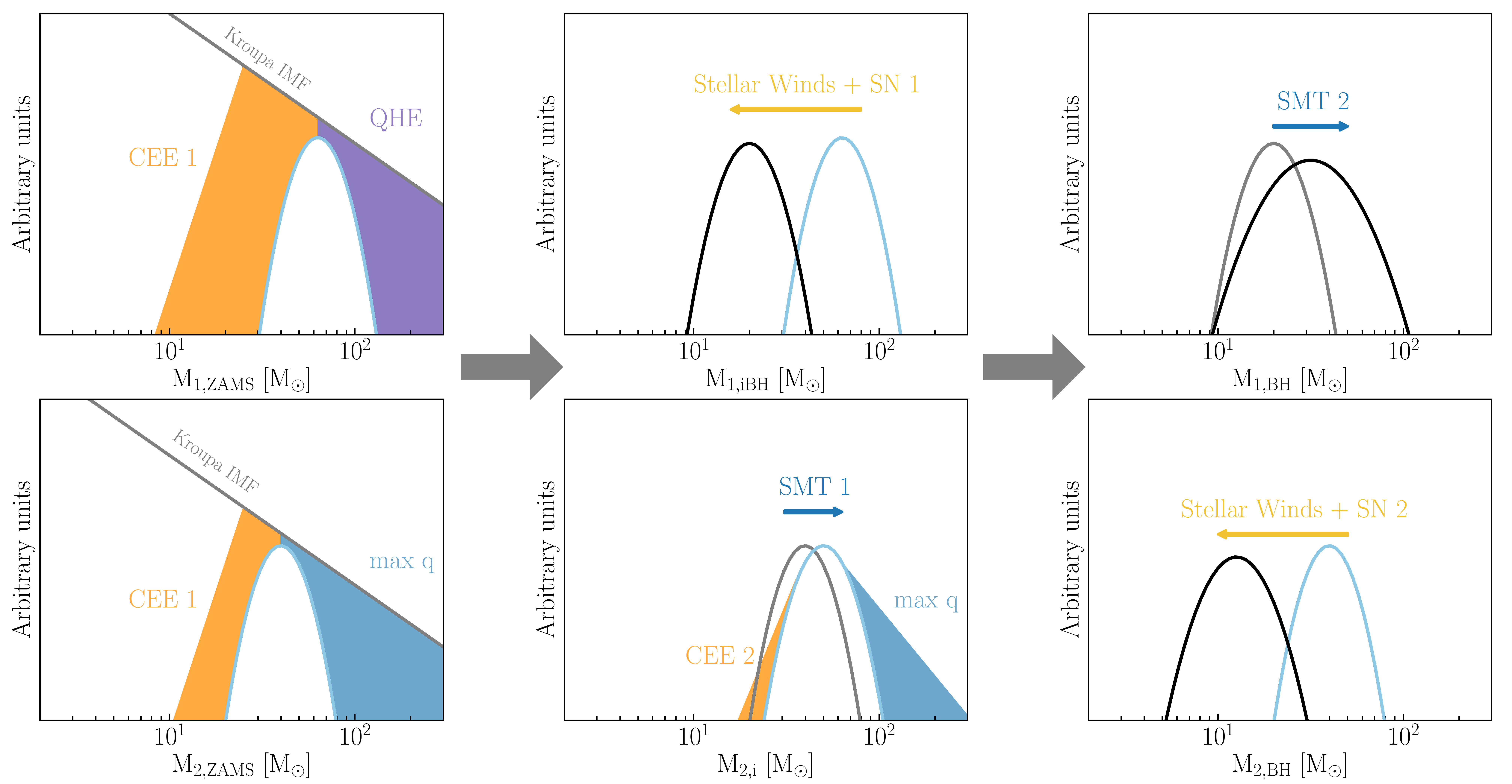}
    \caption{A cartoon depiction of the BBH distribution and the stellar physics that restricts stable mass transfer at low metallicity. The peaks represent the SMT channel at each stage. From the Kroupa IMF (grey), most systems do not result into a BH or merge within the Hubble time. In the primary models, the low ends of the $\MpZAMS$ and $\MsZAMS$ distributions are limited by mass transfer stability (CEE 1; orange), while the high-mass regime is limited by QHE (purple). As a result of limited high-mass contribution, the secondary is also limited (max q; blue). As metallicity increases, the $\MpiBH$ shifts towards lower masses due to stellar winds. Also in the primary model, $\MsZAMS$ increases due to SMT onto it, resulting in the $\Msi$ distribution. In the secondary model, CE restricts $\Msi$ at the low end (CEE 2; orange), while the low end is limited by the maximum mass ratio. During this phase, $\MpiBH$ increases in mass due to SMT to give the $\MpBH$ distribution in the top right Figure. $\Msi$ losses material due to SMT, stellar winds, and the supernova, and results in the $\MsBH$ distribution in the bottom right Figure. Black and blue peaks lines peaks indicate the BH and stellar distribution, respectively. Gray distributions are the distributions from the previous phase.}
    \label{fig:cartoon_QHE}
\end{figure*}

Instead, limitations from mass transfer stability, QHE, and stellar winds restrict the SMT+SMT channel to form BHs around $35 \Msun$. We distinguish between the regime with QHE ($Z < 0.006$) and without, because QHE restricts further interactions in \textsc{bpass}.
Figure \ref{fig:cartoon_QHE} shows a cartoon depiction of how each process in the QHE regime restricts the SMT+SMT formation pathway. Most low ZAMS primary and secondary masses will not form a BH or merge within the Hubble time.

At QHE metallicities, $\MpZAMS$ is restricted at the high end due to QHE (see Section \ref{sec:QHE_limit} for more details). Due to the high masses, these models often interact stably and have a lot of material available for mass transfer. Since the timescales of the primary and secondary as similar (see Appendix \ref{app:homology}), accretion is efficient and the models are likely to undergo QHE. This restricts further interactions and makes it impossible for the QHE systems to merge within the Hubble time. As a result of the limited $\MpZAMS$, $\MsZAMS$ is also restricted, since we define $\MpZAMS > \MsZAMS$.

The low end of $\MpZAMS$ is limited by CE. If the donor star is less massive, it is more likely to undergo unstable mass transfer and, thus, does not contribute to the SMT channel (see Section \ref{sec:MT_stability} about the stability). Similarly if the companion star is too small compared to the primary mass, it will experience CE due to the large mass ratio and restricts the bottom of the secondary mass distribution.

Further limitations are introduced at later stages of the evolution. For example, the formation of the first BH, $\MpiBH$, at the lowest metallicities is limited by the (P)PISN, creating a pile-up around $40 \Msun$. However, the contribution of these events is limited and at higher metallicities stellar winds limit the maximum $\MpiBH$ mass. Furthermore, since the accretion before the first SN is limited due to QHE, the secondary mass distribution ($\Msi$) is similar to the ZAMS distribution ($\MsZAMS$). The exception is that the low end is further restricted by the mass transfer stability of the interaction in the secondary model, e.g. low secondary mass stars are more likely to experience CE, while higher mass systems are more likely to interact stably. At low metallicity this results in mass ratios close to unity. As a result of these mechanism, low metallicity systems are at the PISN limit and have limited accretion. 

As metallicity increases, mass transfer becomes more stable during the core-helium burning. Since the material available for mass transfer during this phase is limited, QHE can be avoided, while $\Msi$ can still transfer a significant amount of material onto $\MpiBH$. We discuss this further in Section \ref{sec:extended_masses}.

At metallicities above $0.004$, QHE no longer restricts the maximum $\MpZAMS$. Instead it is restricted by the IMF, and higher mass primaries and secondaries can interact stably. This increases the systems contributing to the merger rate and the material available for the formation of BHs. At the same time, mass loss due to stellar winds has also increased, which decreases $\MpiBH$ and $\MsBH$. The interplay between the mass transfer and stellar winds results an initial increase in BH mass with increased metallicity till the stellar winds become stronger and reduce the BH mass again. Since mass is stored on the companion and transferred back onto the BH, a transition to BBH systems with high $\MpBH$ and small $\MsBH$ (q closer to 0) takes place, as metallicity increases.

\begin{figure*}
    \centering
    \includegraphics[width=\textwidth]{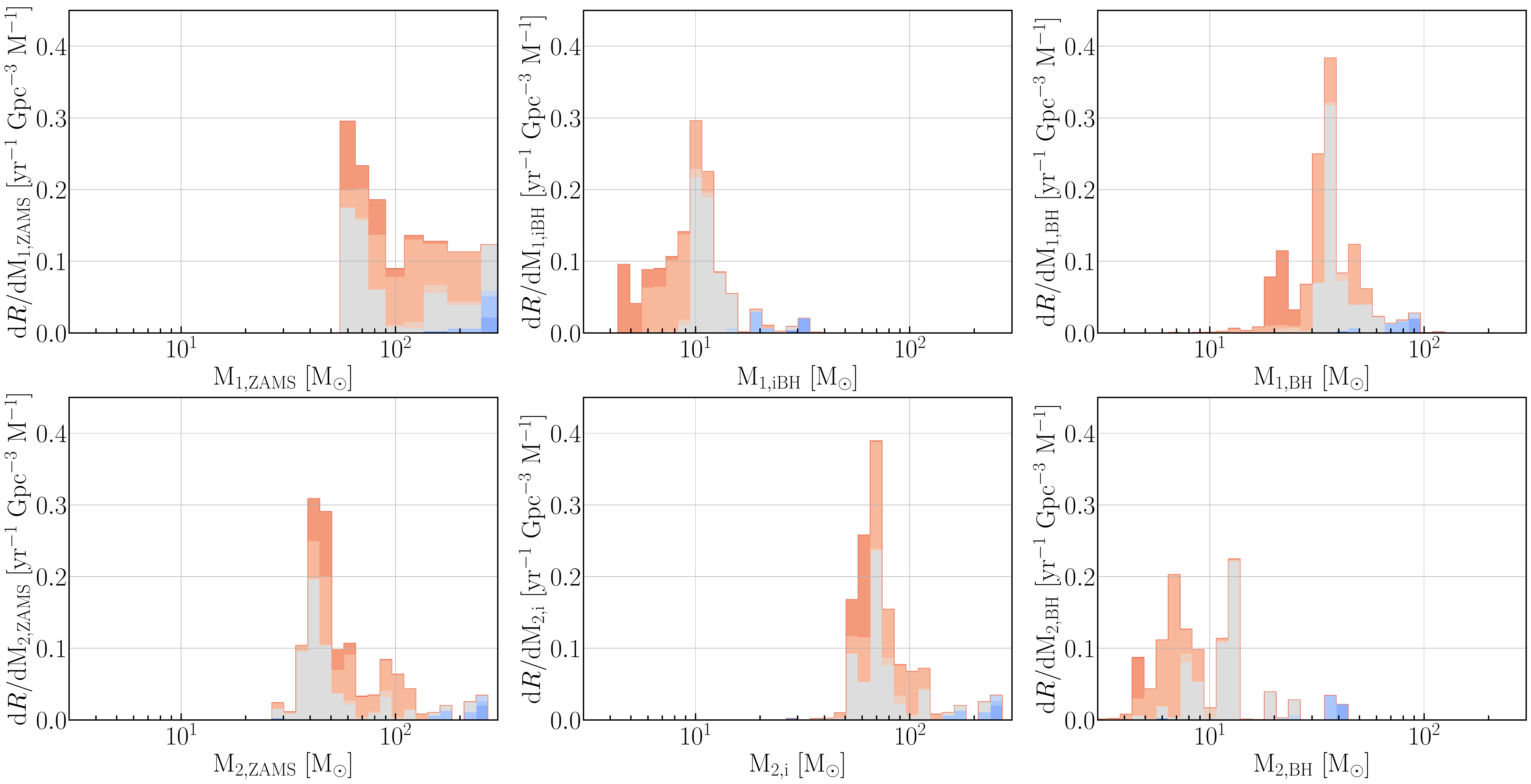}
    \caption{The evolution of the primary and secondary masses in the SMT+SMT channel separated by metallicity. Standard binning (40 bins) is used, except for $\MpZAMS$ for which we use the values of the \textsc{bpass} grid. The left most figures show the ZAMS distribution, the middle figures the masses right after the first supernova, and the right figures the final remnant masses of the merging BBHs in this channel.}
    \label{fig:SMT+SMT}
\end{figure*}

We now turn our attention to the actual distributions in the SMT+SMT channel in Figure \ref{fig:SMT+SMT}. The left most Figures shows that most of the systems originate from $Z>0.004$; above the QHE limit. While the contribution of systems below the QHE limit to the peak is limited, these systems play an important role in the high-end of $\MpBH$ and will be discussed in Section \ref{sec:extended_masses}.

Above the QHE limit, systems with $Z=0.006$ (grey) cover a large range of $50 < \MpZAMS < 300$. As metallicity increases (redder), the high end of $\MpZAMS$ becomes restricted due to stellar winds keeping the radius of the star small and limiting interactions. Furthermore, the material available for mass transfer and $\MpiBH$ formation reduces. Thus, as metallicity increases $\MpiBH$ decreases and the change in mass between $\MsZAMS$ and $\Msi$ decreases.

The $\MpiBH$ distribution after the first SN is similar to the single star remnants with the same ZAMS mass, but is more extended to the low and high end. The less massive BHs are due to enhanced mass loss, reducing the core mass of stars and leading to less massive remnants as discussed by \citet{laplace_2021}. The more massive BHs are, however, more surprising, but were also found in \citet{eldridge_2016}. These are a consequence of mass transfer preventing the opening of a convective zone at the edge of the helium core post-main sequence. This convective region would normally dredge-up core material decreasing its mass. By preventing its formation, the final core mass is greater than expected. This process is only revealed by the detailed evolution models.

With limited material available for the BH to grow during the second phase of SMT, $\MpBH$ decreases with increasing metallicity. Due to this process, the $35 \Msun$ peak contains a metallicity distribution, where as $\MpBH$ increases, metallicity decreases. The combination of stellar wind, QHE, and stability criteria result in the placement of this peak near $35 \Msun$. Furthermore, $\Msi$ has additional mass loss at higher metallicity due to stellar winds. Due to which, $\MsBH$ decreases and the mass ratio of the final BBH systems increases. Moreover, these systems are also formed more recently and most require eccentric orbits to be able to merge within the Hubble time.

\subsection{The formation of upper mass gap BHs ($> 50 \Msun$)}
\label{sec:extended_masses}

The formation of high-mass BH is difficult due to the PPISN and PISN mechanism reducing the BH progenitors mass or completely disrupting it, respectively. Furthermore, the formation of a common envelope will shed a significant amount of stellar material from the system, further reducing the available mass for BH formation. In agreement with \citet{neijssel_2019} and \citet{vanson_2022}, we find that high-mass BH ($20 - 115 \Msun$) are increasingly formed through systems only undergoing stable mass transfer, as shown in Figure \ref{fig:m1_q_distributions}. However, we find significantly higher masses than predicted by rapid population synthesis codes \citep{bavera_2021b, olejak_2021,vanson_2022}.

This is a direct result of super-Eddington accretion onto the BH, which results in no mass loss from the system during accretion, except for wind drive mass loss. However, \citet{vanson_2020} and \citet{bavera_2021b} have shown that a super-Eddington accretion rate onto a BH leads to fewer mergers within the Hubble time due to reduced mass and angular momentum loss from the system. Thus, for the systems to be able to merge, their periods need to shrink sufficiently during the systems evolution. This can either happen through angular momentum loss from the system, mass transfer from the more massive star to its companion, or tidal synchronisation. Since super-Eddington accretion limits the former, we focus our attention on the latter two.

The $\MpBH$ distribution in Figure \ref{fig:formation_channels_met} shows that the high-mass BHs come from high-mass sub-QHE limit metallicity $\MpZAMS$ systems with massive $\MsZAMS$. During the initial phase of the evolution mass is transferred to the companion, but the amount transferred is limited, because QHE would limit further interactions, if accretion was more than 5 per cent of the companion mass. The formed $\MpiBH$ is, therefore, limited by (P)PISN. As a result of the supernova, the systems have large mass ratios ($q_\text{2,final} \sim 5-10$).  But because $\Msi$ are massive, they are able to interact stably (see Section \ref{sec:MT_stability}) and due to the interaction taking place on the main-sequence, a significant amount of material is transferred onto the BH.

Since the initial mass ratio was larger, mass is transferred from the most massive star to the less massive BH, which reduces the period. Furthermore, as a result of the high mass ratio, tidal forces are strong and further reduces the period. A CE is avoided by the 'fast' radial shrinkage of the donor star, and the 'fast' increase of the BH mass due to the super-Eddington accretion, as such the mass ratio approaches unity and the radius of the donor star does not reach the separation of the system. These interactions are on a longer than thermal timescale, although the thermal timescale is short due to the massive nature of the donor star.


\section{Mass Transfer Stability} \label{sec:MT_stability}
We have shown that using detailed stellar models with PPISN and super-Eddington accretion onto BHs leads to an extended mass tail up to $100 \Msun$ and an overdensity near $35\Msun$, as observed by the LIGO/VIRGO/KAGRA collaboration. These are both an effect of increased stable mass transfer. As such, this section covers an overview of the nature of the detailed models, how the stability criteria influence the features in the $\MpBH$ distribution, and a comparison against more detailed mass transfer stability determination. 
The Supplementary Material explores the stability of the primary and secondary models in \textsc{bpass}.

\subsection{Nature of the envelope}
As discussed in Section \ref{sec:MT}, a star can respond to mass loss by expanding or contracting, which depends on the properties of the envelope of the star. In general, if a star as a convective envelope it expands due to adiabatic mass loss, while it contracts if the envelope is radiative \citep{hjellming_1987, soberman_1997}. However, work by \citet{ge_2010, ge_2015, ge_2020, ge_2020a} has shown that metallicity, radius and evolutionary phase also influence the mass transfer stability. Most rapid population synthesis BBH merger predictions do not take this into account \citep[\textsc{startrack}, \textsc{mobse}, \textsc{cosmic}, \textsc{compas};][]{belczynski_2016, giacobbo_2018a, breivik_2020, riley_2022} and instead use an adiabatic model based on the evolutionary phase to determine the stability of mass transfer, as per \citet{hurley_2002}. This leads to more unstable mass transfer, since, for example, convective envelope criteria are applied to core-helium burning stars that can have radiative envelopes \citep[for an overview see][]{klencki_2020, klencki_2021}. Furthermore, other detailed mass transfer simulations have shown that BH-star systems are more likely to undergo SMT \citep{pavlovskii_2017, marchant_2020, marchant_2021}, impact population of merging BBHs \citep{gallegos-garcia_2021}, and that the donor response is very different than the simplified adiabatic model and holds across stellar codes (\textsc{stars}, \textsc{mesa}, Heyney-type code) \citep{woods_2011, passy_2012}.

\begin{figure}
    \centering
    \includegraphics[width=\columnwidth]{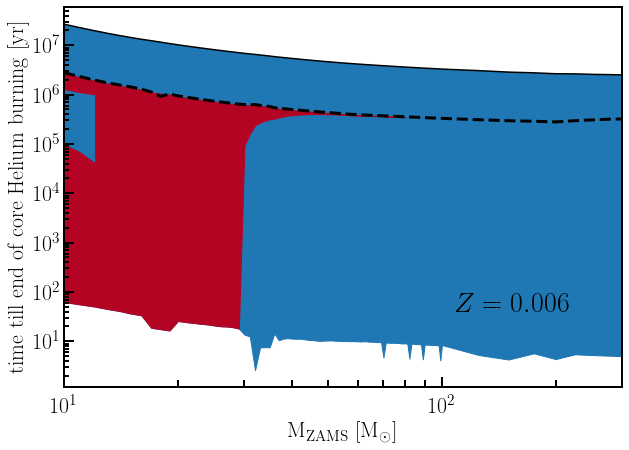}
    \caption{The radiative (blue) or convective (red) nature of the envelope of single stars at $Z=0.006$ from ZAMS to the end of Helium core burning. The process for extraction from the single star models is discussed in the Supplementary material. For models that did not reach the end of core helium burning, we interpolate between ZAMS masses. We indicate the ZAMS (solid black line) and the end of the Main-Sequence (dashed black line).}
    \label{fig:nature_z006}
\end{figure}

Since \textsc{bpass} is based on the Cambridge \textsc{stars} code, stability is determined by following the equations of stellar structure through mass loss, which allows us to determine the nature of the envelope over a large mass and metallicity range. Since mass transfer alters the stellar evolution and the internal structure of the donor star, we turn our attention to single stars to determine the convective or radiative nature of the donor envelope, because this is the structure of the star just before the onset of RLOF. Section 1 in the Supplementary Material covers the evolution of the envelope over age, mass, and metallicity. We find that at nearly all massive stars spend their main-sequence with a radiative envelope. Post-Main Sequence the envelope depends on the initial mass and metallicity. At $Z>0.002$, $\textrm{M}_\textrm{ZAMS} \leq 26$ have convective envelopes after core-helium burning initiates. Figure \ref{fig:nature_z006} shows this for $Z=0.006$. Above this mass, the envelopes only have a short period of convection before core-helium burning initiates, after which the star has a radiative envelope, which is similar to results found by \citet{klencki_2021}. This indicates that interactions during this phase are more stable than estimated by rapid population synthesis codes. Below $Z < 0.002$, the star becomes convective during core helium burning at $\textrm{M}_\textrm{ZAMS} \geq 21$. Below this limit, the absence of metals restricts the formation of a convective envelope till late in the core-helium burning phase. At $Z=10^{-5}$, the convective zone is completely avoided at these low masses.

This shows that metallicity, age, and mass all influence the nature of the envelope in single star. In binaries, short mass transfer phases could alter the internal structure of the star in such a way that at later stages in the donors evolution the envelope no longer becomes convective. Thus, without detailed treatment of the internal structure at the moment of mass, BHs formed through SMT are missed in rapid population synthesis codes.

\subsection{Mass Ratio Exploration} \label{sec:q_explosation}

The resulting $q$ values for SMT onto the BH in Figure \ref{fig:q_SMTvsCEE} are more extreme than typically found in detailed binary models \citep{marchant_2021, gallegos-garcia_2021}. Although the mass ratios at the moment of mass transfer are less extreme, values up to $q=23$ remain. This could indicate that our determination for CE is too constricting, since our stability determination only considers the donor radius and the separation of the system. 
Thus, in this section, we explore how the high-mass features in $\MpBH$ depend on the mass ratios and explore the extreme mass ratio systems.

\subsubsection{Mass Ratio Cuts}
\begin{figure}
    \centering
    \includegraphics[width=\columnwidth]{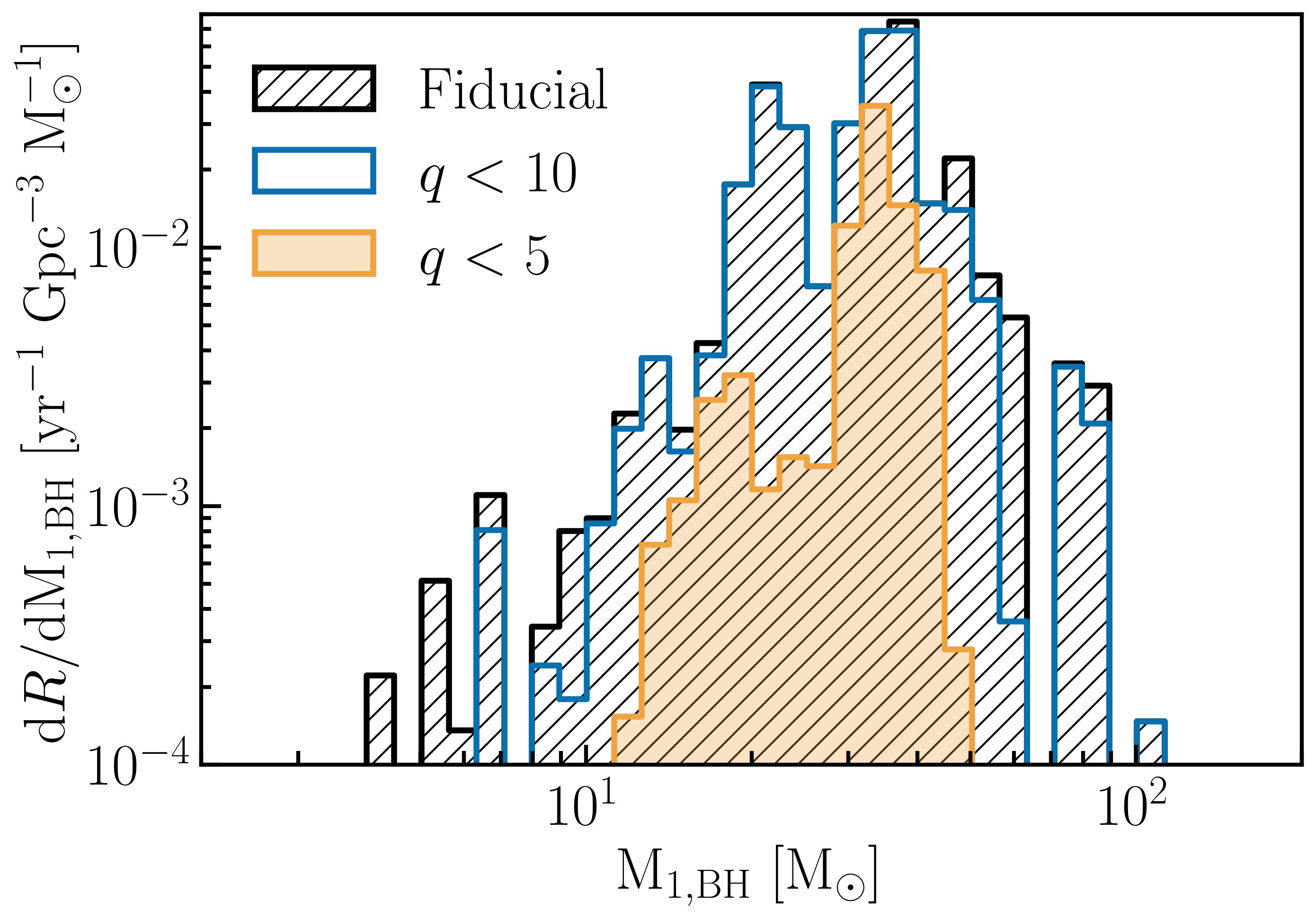}
    \caption{The primary BH mass ($\MpBH$) distribution at $z=0$ from the SMT+SMT channel with the fiducial model (black) and two cuts based on the mass ratios at the moment of mass transfer with $q<10$ (blue) and $q<5$ (orange).}
    \label{fig:q_limit}
\end{figure}

Since the high-mass features predominantly come from the SMT+SMT channel, we perform two cuts at $q<10$ and $q<5$ based on the mass ratio at the moment of mass transfer onto the BH. The $q<10$ cut in Figure \ref{fig:q_limit} shows that large mass ratios ($q>10$) are only a fraction of the SMT+SMT channel. Moreover, the cuts show that the $35 \Msun$ peak is dominated by $q<5$ systems with the spread around the peak being a result of more extreme mass ratios ($q<10$). As a consequence, the upper mass gap systems undergo SMT with mass ratios between 5 and 10, and the stability criteria could influence the existence of these systems. These mass ratios are larger than generally considered stable, but are not unreasonable at low metallicity and high donor mass. In Section \ref{sec:comparison}, we compare our stability criteria other detailed work, and the Supplementary Material describes that these models are stable due the super-Eddington accretion and faster shrinkage of the donor star radius than the Roche Lobe.

\subsubsection{Extreme Mass Ratios}
Some BH-star systems undergoing have extreme mass ratios with $q_\text{2,initial}$ up to 25, as shown in Figure \ref{fig:q_SMTvsCEE}. The actual mass ratio at the moment of RLOF is less extreme due to mass loss, but some $q>15$ remain. These systems all interact on the main-sequence and occur between a small BH (3-6 $\Msun$) and a large star ($70\textrm{--}140 \Msun$) at $Z>0.01$, where stellar winds are strong. As a result, the primary loses mass quickly, while the amount of RLOF is small. This reduces the mass ratio and, thus, avoids the Darwin Instability. Due to the strong stripping of the star, stellar winds become stronger and limit further interactions. Although these systems exist, their contribution to the $\MpBH$ distribution is minimal and limited to the low mass regime.

\subsection{Comparison to the literature} \label{sec:comparison}

Because some $q\approx25$ systems undergo SMT and the $q \approx 5\textrm{--}10$ for the high-mass $\MpBH$ is higher than conventional main-sequence interactions, we compare the \textsc{bpass} stability to other detailed analyses of the stability of mass transfer. In the Supplementary Material, we further describe the mass transfer stability in \textsc{bpass}.

\begin{figure}
    \centering
    \includegraphics[width=\columnwidth]{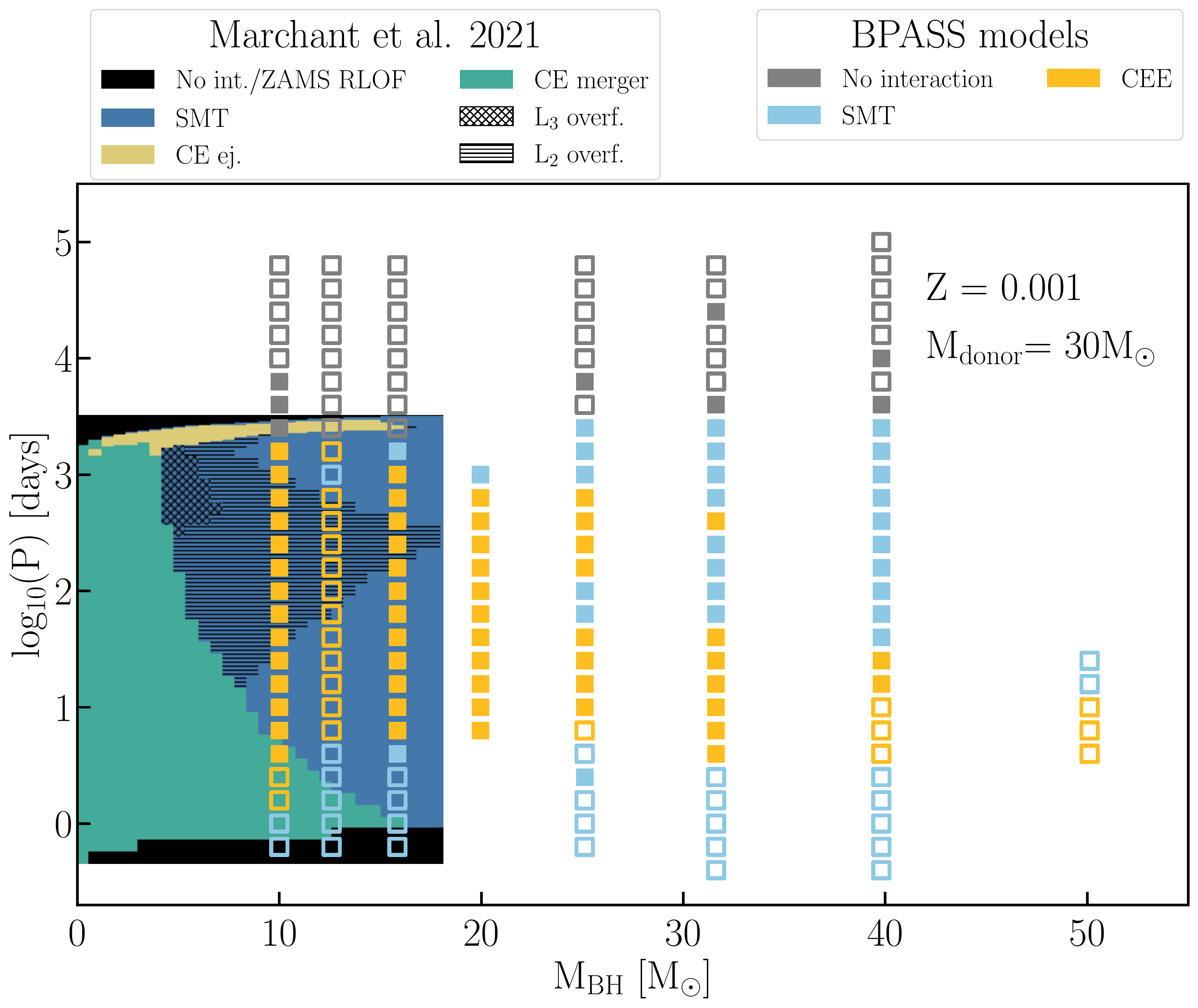}
    \caption{Mass transfer stability comparison between \citet{marchant_2021} models ($Z=0.00142$) and the \textsc{BPASS} models ($Z=0.001$). The \citet{marchant_2021} CE and SMT interactions leading to BBH mergers within the Hubble time have been grouped together to those taking longer. The solid squares mark the models contributing at the event rate at $z=0$, while the open squares do not contribute to the BBH merger rate.}
    \label{fig:marchant}
\end{figure}

\citet{marchant_2021} explored the stability of mass transfer using detailed treatment of the mass flow through $\textrm{L}_1$ and mass loss through the outer Lagrangian point of the donor star ($\textrm{L}_2$/$\textrm{L}_3$) using \textsc{mesa} \citep{paxton_2011, paxton_2013, paxton_2015, paxton_2018, paxton_2019} at a metallicity of $Z=0.00142$ and a $\Mdonor =30 \Msun$. Figure \ref{fig:marchant} shows their results with overlapped the stability of the \textsc{bpass} models at the closest metallicity ($Z=0.001$) at $M_\textrm{donor} = 30\Msun$.

Although the overlap between the simulations is limited, \textsc{bpass} contains CE in regions where SMT is predicted by \citet{marchant_2021}. This is most likely due to the synchronisation of the donor spin with the orbit when RLOF takes place, which shrinks the orbital separation and in general leads to more frequent CE, which is not considered in \citet{marchant_2021}. Interestingly, non of the CE interaction at this metallicity and mass in \textsc{bpass} lead to a merger, but they do not probe far into the CE merger regime from \citet{marchant_2021}.

In \textsc{bpass} SMT occurs pro-dominantly during the main-sequence and during core-helium burning, when the envelope of the donor star is radiative and the star evolves on a nuclear timescale. When the mass ratio is reversed ($\Mdonor \lesssim \mathrm{M}_\mathrm{BH}$), the system can interact stably during the Hertzsprung gap at $Z=0.001$ and donor mass. However, the stability during this phase is highly dependent on the donor mass, metallicity, and mass ratio (see Supplementary Material).

A larger donor mass range using similar stellar evolution was explored by \citet{gallegos-garcia_2021} at $Z=0.00142$. As with \citet{marchant_2021}, we find that more CE takes place due to the tidal forces, and that more SMT takes place on the main-sequence. Furthermore, similar to \citet{gallegos-garcia_2021}, we find that for higher donor masses, the stability on the main-sequence increases.


While the above comparison indicates that \textsc{bpass} has more CE at the donor masses explored by \citet{marchant_2021} and \citet{gallegos-garcia_2021}, the majority of companion masses in the SMT+SMT channel have masses $\gtrsim 30 \Msun$ and are at a metallicity higher than $Z=0.001$, see Figure \ref{fig:formation_channels_met}.
\citet{ge_2020} explores thermal timescale mass transfer and $L_2$ overflow at an even larger range of donor masses and at a higher metallicity ($Z=0.02$). Since nearly all of our interaction take place on a longer than thermal timescale, we look at the critical mass ratio from $L_2$ overflow (their figure 9). All the donor stars in our secondary models, that lead to BBH mergers at $z=0$, lie on the right side of their figure 9. In general, we find that if the star fills its Roche Lobe, while evolving on a thermal timescale, such during the Hertzsprung gap, the system undergoes CE. The \textsc{bpass} models interact stably during their Main-Sequence evolution and in the late stages with a few exceptions. 

When the star interacts with the BH during the Main-Sequence, the critical mass ratio is less than in later stages of the evolution according to \citet{ge_2020}. However, it is important to note that the stability criteria are at $Z=0.02$, while the SMT in the secondary model in \textsc{bpass} comes from lower metallicities, especially when considering the upper mass gap BHs. At lower metallicities, the opacity and, thus, radius is small, as such the $L_2$ overflow could take place at higher mass ratio on the Main-Sequence. However, a more detailed analysis of the interactions is required.

The closest matching BH and donor mass system for which stability has been explored is by \citet{shao_2022}, who explored case A interactions between a $30 \Msun$ BH and a $80 \Msun$ donor star at $Z = 0.02$ and $Z=0.001$ using \textsc{mesa} with limited super-Eddington accretion ($10\text{M}_\text{Edd}$). These are similar to the \textsc{bpass} models that are progenitors to the PISN mass gap BHs. 

\textsc{bpass} does not treat outflow out of the outer Lagrangian point, which could result in more CE and mergers, but also the opportunity for more SMT systems to reach periods that can merge within the Hubble time. However, in \textsc{bpass} the CE that occur just Post-Main Sequence early in the HG appear very different to those that occur with a deep convective envelope. In the former the orbit radius changes little, while in the latter significant orbit shrinkage occurs. In the former this would be similar to mass loss through the $L_2$ point with not much mass transfer occur and the orbit remaining constant compared to the strong CE we expect at higher periods and separations.
Furthermore, whether or not the star expands past its outer Lagrangian point, is determined by the mass loss rate due to mass transfer. \citet{marchant_2021} showed that a more detailed prescription of this rate could increase the mass loss and reduce the expansion of the star, which limits the $L_2$ outflow.
However, the interplay between mass loss rate due to mass transfer and outflow through the $L_2$ point impacts the evolution of the systems in a non-linearly and has not been explored at the high donor masses and metallicities that form the highmass regime of the $\MpBH$ distribution. A more detailed investigation is required to understand the influences of the mass transfer rate at high donor star masses at low metallicity and their impact on the observed BBH merger rate, but this is beyond the scope of this work.

\section{Robustness of $\MpBH$ features} \label{sec:robustness}

As discussed in Section \ref{sec:primary_mass_features}, many aspects of stellar evolution come together to shape the features the $\MpBH$ distribution. The BBH rate and distribution is rather robust against some evolutionary parameters, such as the natal kick prescription \citep{broekgaarden_2021a}. This is further confirmed by Figure A3 in \citet{ghodla_2022}, where the $35 \Msun$ overdensity and extended tail remain between different natal kick prescriptions. However, the BBH rate and distribution is very dependent on the star formation history and metallicity evolution \citep{ chruslinska_2019, tang_2020, broekgaarden_2021a}. In \citet{briel_2022}, we have shown that the combination of \textsc{bpass} and the TNG star formation history results in electromagnetic and gravitational wave transient close to observations. Moreover, the features in the high-mass $\MpBH$ distribution remain when using the empirical star formation history from \citet{briel_2022}. The stellar wind prescription can also alter the merging primary mass BH distribution by altering the mass available for mass transfer and the compact remnant \citep{broekgaarden_2021a,dorozsmai_2022}.

Section \ref{sec:MT_stability} covered the influence of the stability criteria on the primary BH mass distribution features. In the following Sections, we explore how the QHE limit and remnant mass prescriptions influence the distribution. 

\subsection{Quasi-Homogeneous Evolution Limit} \label{sec:QHE_limit}

\begin{figure}
    \centering
    \includegraphics[width=\columnwidth]{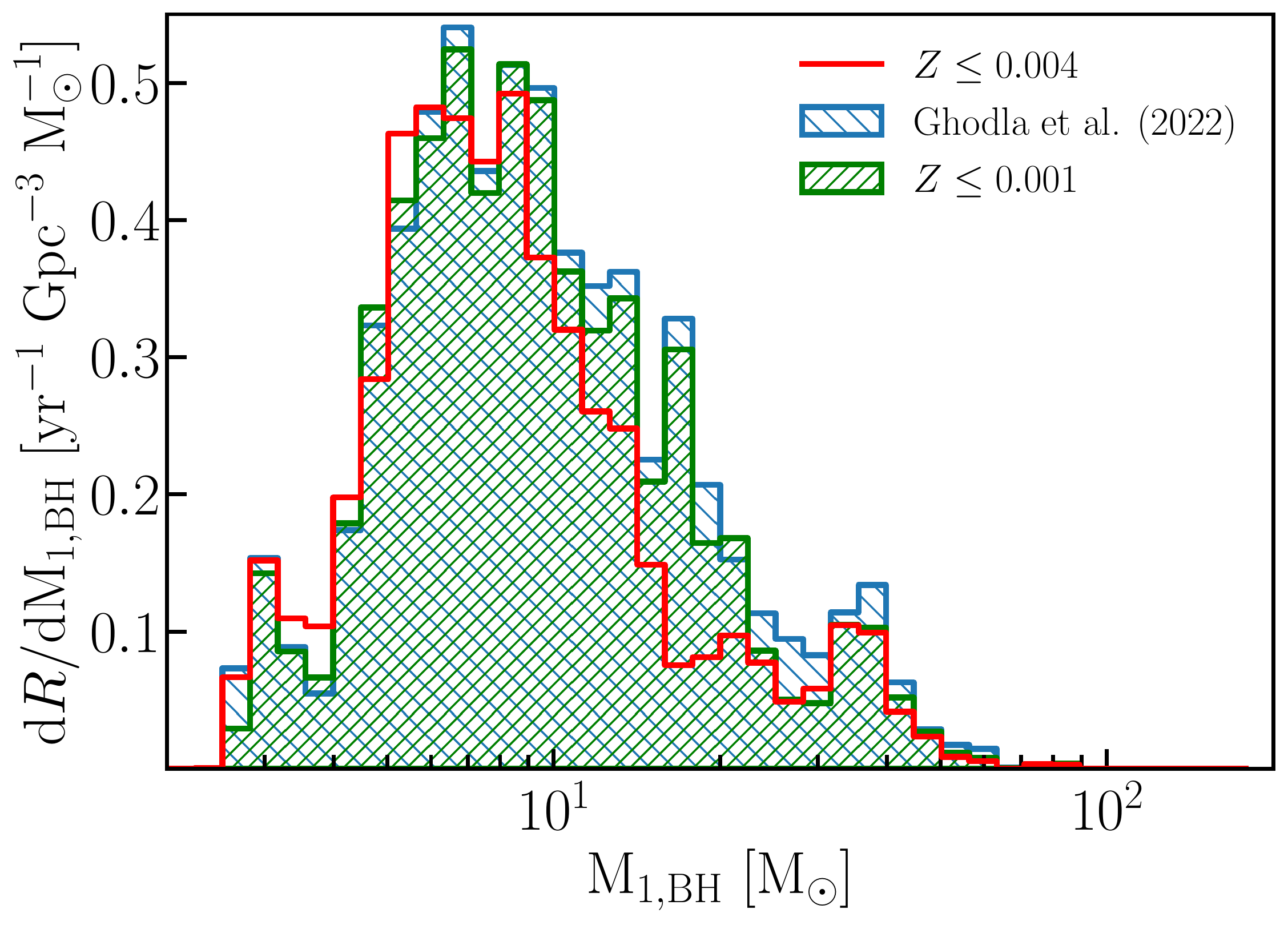}
    \caption{The $\MpBH$ distribution at $z=0$ for the QHE limit ($Z \leq 0.004$) implemented in standard \textsc{bpass} (red), lowered to $Z \leq 0.001$ (green), and the accretion based QHE prescription from \citet{ghodla_2022a} (blue).}
    \label{fig:QHE_limits}
\end{figure}

The fiducial version of \textsc{bpass} uses a hard QHE limit, where below $Z=0.006$ QHE takes place if the companion star accretes more than 5 per cent of its initial mass. Together with the stellar winds, this determines the upper-edge of the $35 \Msun$ excess. As shown in Figure \ref{fig:QHE_limits}, lowering the QHE limit from $Z \leq 0.004$ to $Z \leq 0.001$ causes an increase in high-mass $\MpBH$ around $40 \Msun$ and around the $10 \Msun$ peak, while pushing it towards higher BH masses. Furthermore, a plateau around $20 \Msun$ becomes clear as a result of SMT. While QHE is likely to occur at metallicities $0.001 \leq Z \leq 0.004$ , this shows that QHE restricts the formation pathway for BBH in \textsc{bpass} by limiting further binary interactions. Although this choice is physically motivated, in some cases, the star might spin down and still expand resulting in binary interactions.

Instead of a hard limit, QHE through accretion is more likely to be a more gradual process dependent on the mass and angular momentum accreted. Using a more detailed prescription for accretion QHE, created using \textsc{mesa}, \citet{ghodla_2022a} has shown that in \textsc{bass} less systems undergo QHE, especially at low ZAMS masses. In the high ZAMS mass regime, QHE remains similar to the fiducial \textsc{bpass} model. In Figure \ref{fig:QHE_limits}, we show the primary remnant mass distribution using the QHE prescription from \citet{ghodla_2022a}. The more detailed prescription adds additional systems to the high-ends of the $10 \Msun$ peak and the $35 \Msun$ excess, shifting both to a slightly higher mass, while the upper mass gap BHs remain.

\subsection{Remnant mass prescriptions} \label{subsec:remnantmass}

While we are able to accurately predict the high end of the primary BH mass distribution, the lower end of our prediction does not align with observations. Since this regime is dominated by common envelope evolution, our prescription may need adjustment, but this is beyond the scope of this work.
Another option could be the remnant mass prescription, since this will mostly influence the lower mass regime, except for the (P)PISN prescriptions which influences the high mass regime. To explore the former, we implement the rapid and delayed remnant mass prescriptions of \citet{fryer_2012} in Figure \ref{fig:rem_prescriptions}. And we alter the (P)PISN prescriptions to show that the high mass features in the $\MpBH$ distribution remain.

\subsubsection{Fryer Rapid/Delayed} \label{subsubsec:fryer}
\begin{figure*}
    \centering
    \includegraphics[width=\textwidth]{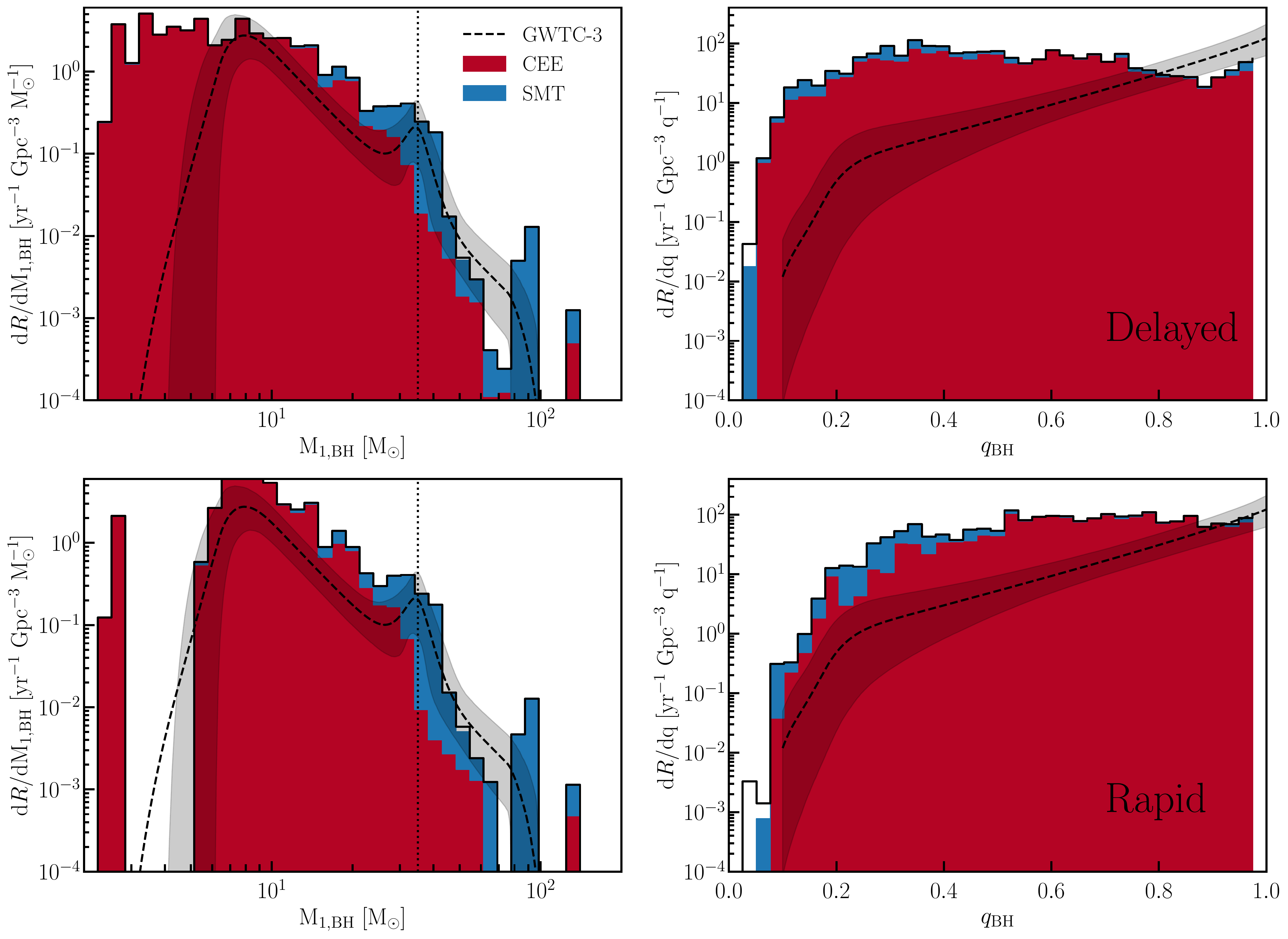}
    \caption{The primary BH mass (left) and mass ratio (right) distribution at $z=0$ of the \citet{fryer_2012} delayed (top) and rapid (bottom) remnant mass prescriptions, split into CEE and SMT formation channels. The Rapid prescription has some BHs with small primary masses, possibly due to accretion onto a NS, such that a BH is formed. $35 \Msun$ is marked with a dashed line.}
    \label{fig:rem_prescriptions}
\end{figure*}

To explore the effect of the remnant mass prescription, we have implemented the rapid and delayed remnant mass prescription of \citet{fryer_2012} in Figure \ref{fig:rem_prescriptions}. Note that these do not implement PPISN.

Both prescriptions increase the number of systems around $10 \Msun$, bringing the predicted distribution above the observed intrinsic rate. This is most likely a result of the fact that fiducial \textsc{bpass} injects $10^{51}$ erg into the star, which could cause too much material to be ejected from low mass stars. However, the Delayed prescription also increases the number of systems between 2 and 5 $\Msun$, while Rapid does not include these systems and has a mass gap between 3 and $5 \Msun$. Figure \ref{fig:CO_core_vs_rem} shows that the Rapid and Delayed prescription are similar in the high mass regime, since both assume full fallback onto the BH. In the low mass regime, the prescriptions differ significantly from each other and from the \textsc{bpass} prescription. \textsc{bpass} generally predicts smaller remnant masses for larger CO cores than both Fryer prescriptions do, and could result in the significantly different low $\MpBH$ regime. While this makes it difficult to untangle the influence from CEE, it also shows that the extended tail and excess near $35 \Msun$ remain with other remnant prescriptions than implemented in fiducial \textsc{bpass}.



\subsubsection{(P)PISN}

We have implemented the PPISN prescription from \citet{farmer_2019} for $\text{M}_\text{CO} \geq 38$, which results in a smooth transition between different remnant mass prescriptions. \citet{vanson_2020} found an additional bump at $30 \Msun$ as a result of a non-smooth transition between the CCSN and PPISN prescription. This is not present in our remnant mass distribution. Moreover, fiducial \textsc{bpass} does not contain this PPISN prescription, but the $35 \Msun$ excess is still present.
\begin{figure}
    \centering
    \includegraphics[width=\columnwidth]{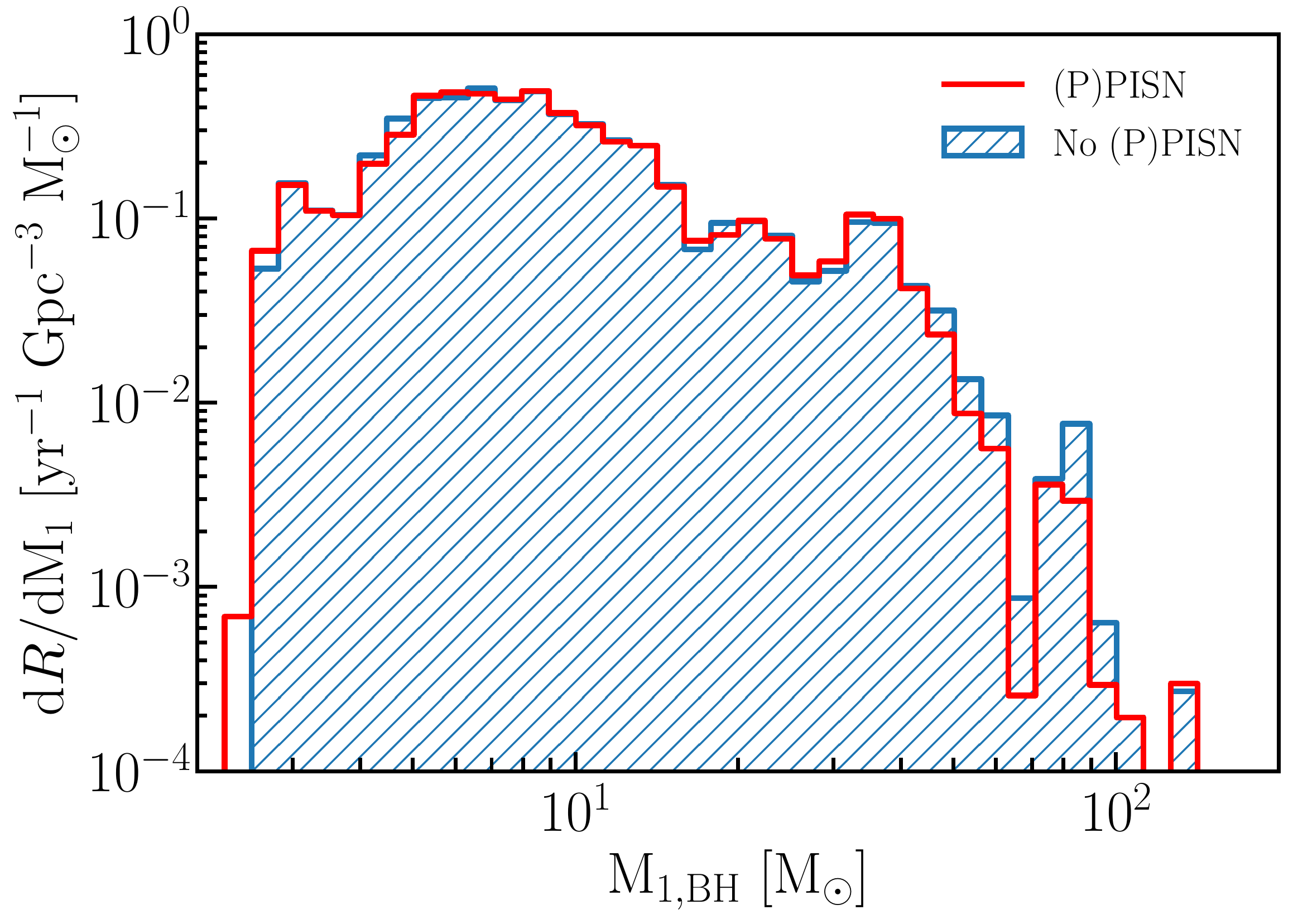}
    \caption{The $\MpBH$ distribution at $z=0$ of merging BBHs for the model presented in this work with PPISN and PISN (red) and for a model without PPISN and PISN (blue). The impact of (P)PISN in \textsc{bpass} on the distribution is minimal.}
    \label{fig:PISN}
\end{figure}

Furthermore, in Figure \ref{fig:PISN} we have removed the PPISN and PISN prescriptions from the population synthesis and find only subtle changes in the merger rate in the PISN mass gap region. This in contrast to findings by \citet{stevenson_2019}, who found that the (P)PISN influences the high-mass $\MpBH$ regime. Because most of the high primary mass systems in \textsc{bpass} are created through accretion onto BHs with initial masses between $10-30 \Msun$, the impact of (P)PISN is limited in \textsc{bpass}, as those BHs do not come from progenitors experiencing PPISN.

\section{Caveats and Uncertainties} \label{sec:open_problems}

\subsection{BH spin} \label{sec:BHspin}

Besides stability of mass transfer, the efficiency of mass transfer is an important factor. To achieve high-mass BHs, we require stable mass transfer with super-Eddington accretion onto the BH. This can leave an imprint on the spin of the population, since a large amount of material is accreted from the companion. This should lead to a non-negligible spin \citep{zevin_2022} and even high spins for case A stable mass transfer \citep{shao_2022}, albeit it is unclear if the BH remains rotating \citep{tchekhovskoy_2012}.
\textsc{bpass} does not currently track the spin of BHs. Thus, as an alternative, we look at the amount of accreted material onto the BH, which we use as a proxy for the spin. 


We find that most material is accreted by BHs in systems with $q \lesssim 0.5$, which would results in these BHs spinning with a positive spin. Furthermore, high chirp-mass systems have a lot of accreted material, as is expected, since the large chirp mass systems require material to remain in the systems. These relations are in agreement with relations for SMT systems found by \citet{zevin_2021, zevin_2022}, and could align the observed correlation between high chirp-masses ($\mathcal{M} > 40 M_\odot$) and a positive effective spin found by \citep{abbott_2021c} under the assumption of thin disk accretion \citep{bardeen_1970, king_1999}. As such, further analysis is required to find the spin of our predicted merging BBH systems.

Another observable could be the observed BH+star binaries, which are often quickly rotating, with BH masses around $10\textrm{--}20 \Msun$ and have high companion masses ($31-70 \Msun$) \citep[for an overview of systems, see Conclusion of][]{shao_2022}. These systems might have been spun up during a previous mass transfer phase before a the current observed mass transfer phase. This is similar to the interactions in our models, where an initial SMT interaction takes place on the Main-Sequence, followed by a SMT or CEE interaction during a later stage of the evolution. The observed companions are often a giant or supergiant and have short periods \citep{miller-jones_2021,orosz_2007, orosz_2009}. Albeit depending on the metallicity and mass ratio, our models undergo SMT if the star is in this phase. However, a detailed analysis of these BH binaries is required to determine their presence in \textsc{bpass} and relation to the BBH merger population.

\subsection{Super-Eddington Accretion}

Super-Eddington accretion is required to conserve enough mass in the binary system to form massive BHs, but also to allow for more stable mass transfer due to a fast changing mass ratios.
For spherically symmetric accretion, the Eddington Luminosity restricts the accretion rate onto a BH. The accretion rates onto the BH in our models are generally more than 100 times Eddington limited accretion and in rare occasions reaches $10^4$ times the Eddington rate. If the accretion is limited, it is uncertain whether or not the BH is still able to accrete similar amounts through longer accretion periods.

A radiation pressure dominated disk could increase the accretion rate up to 10 times Eddington \citep{begelman_2002, ruszkowski_2003}, which can be sufficient to grow the BH. As shown by \citet{mapelli_2009} and \citet{zampieri_2009}, marginal super-Eddington accretion could result in $30-80 \Msun$ BHs from low metallicity environments.

However, most \textsc{bpass} models accrete at a higher rate. To achieve these, other accretion methods have to occur, such as near radial inflow, neutrino emission, advection into the event horizon, or semi-relativistic polar outflows \citep{popham_1999, begelman_2002, ruszkowski_2003,  sadowski_2016, takeo_2020,yoshioka_2022}. These can result in super-Eddington accretion rates between 100 and 1000 times the limit, which where most of the accretion in our models occur. 

Since the super-Eddington accretion takes place on a longer than thermal timescale, these systems spend a reasonable amount of time transferring mass at a super-Eddington rate. This should make it possible to observe these systems, especially because they could emit at a super-Eddington luminosity \citep{klencki_2022}. Observationally, these could be similar to very and ultra luminous X-ray sources, like Holmberg II X-1 \citep{cseh_2014}, M101 X-1 \citep{liu_2013, shen_2015}, M83 ULX-1 \citep{soria_2012, soria_2015}, IC 342 X-1 \citep{das_2021}. These systems are thought to consists of a companion star and a stellar mass BH with super-Eddington accretion \citep{ebisawa_2003, motch_2014, ogawa_2021,  wielgus_2022, ambrosi_2022}, although this is an area of active discussion, since it could also be an intermediate mass BH \citep{ramsey_2006}.

\subsection{Mass Ratio Reversal} \label{subsec:MRR}
Super-Eddington accretion does not only leave an imprint on the spin of the systems, but also the mass ratios. Recent work from \citet{zevin_2021} has indicated that up to 72 per cent of BBH systems can undergo mass ratio reversal in optimal condition and even up to 82 per cent in populations from \citet{broekgaarden_2022}. In this process, the initially more massive star become the less massive BH in the BBH merger. Observationally, mass ratio reversal is thought to be limited \citep{mould_2022}. By implementing super-Eddington accretion onto the BH, this process becomes rare and only 4 per cent of BBH systems reverse mass ratio in our population due to the material being accreted instead of blown away from the system. The mass ratio reversal that does take place in \textsc{bpass} is restricted to the low $\MpBH$ regime with most being a result of CEE. Thus, super-Eddington accretion and more stable mass transfer leads to limited mass ratio reversal in merging BHs.


\subsection{Additional Substructure in the $\MpBH$ distribution} \label{subsec:additional_structure}
With GWTC-3, modest confidence for more substructure in the primary remnant mass distribution was found with a drop in merger rate at $14 \Msun$ \citep{abbott_2021c}. The \textsc{bpass} primary BH mass distribution has a reduced number of events at $18 \Msun$, as is visible in Figure \ref{fig:PPISN_formation_channels}. This structure is caused by a drop in CEE systems, while the SMT channel does not increase yet. This may be caused by SMT not being able to shrink the orbit of the system sufficiently for it to merge within the Hubble time, while a CE phase is avoided by our mass transfer stability criteria. Albeit less clearly, the same substructure remains between different remnant mass prescriptions, as can be seen in Figure \ref{fig:rem_prescriptions}. Interestingly, in Section \ref{sec:QHE_limit} this substructure becomes a plateau when QHE is limited to lower metallicities. Further investigation of this substructure will be required in the future, when the properties of the observed population are more constraint.

\subsection{Other Formation Pathways}

Due to approximations made in \textsc{bpass} models, the companion star will always be a main-sequence star before the first supernova. This limits the ability to probe interactions of double cored systems.

While we are able to predict high primary BH masses using isolated binary evolution, other formation pathways could contribute to the total BBH merger rate, such as dynamical interaction \citep[for an overview, see]{mapelli_2021}. These would leave their own imprint on the population of merging BBH holes, such as random spin alignment \citep{zevin_2021}. However, due to uncertainties in other region of compact object predictions, such as the stellar physics in the isolated binary evolution, care should be taken when constraining formation pathways \citep{broekgaarden_2021a, mandel_2021}. 

In the high primary BH mass regime, hierarchical mergers have been suggested as a possibility for their formation. However, this would lead to an isotropic orientated spin distribution, which is not currently observed \citep{abbott_2021c}.

\section{Conclusions} \label{sec:conclusion}

We have combined \textsc{bpass}, a population synthesis code with detailed stellar models, with the metallicity evolution and star formation history of the TNG-100 simulation. On its own, the fiducial \textsc{bpass} populations can self-consistently reproduce a number of massive star evolutionary characteristics, young and old stellar populations \citep{wofford_2016, eldridge_2009, eldridge_2017, stanway_2018a}, and transients rates \citep{eldridge_2016, eldridge_2019, tang_2020, ghodla_2022, briel_2022}. We implement a PPISN prescription and predict the properties of the merging BBH population.

\begin{enumerate}
    
    \item We find that similar to \citet{neijssel_2019} and \citet{vanson_2022} that high primary BH mass mergers are a result of stable mass transfer. Moreover, our stability determination and implementation of super-Eddington accretion results in an extended primary BH mass distribution up to $100 \Msun$ with an excess at $35 \Msun$.
    
    \item The $35 \Msun$ peak is not dominated by PPISN, but is a result of stable mass transfer, QHE, and stellar winds. While PPISN systems still contribute to this peak, it is not the major formation pathway for these systems. Instead, QHE limits the high-end of the peak, while unstable mass transfer limits the lower edge. In combination with the stellar winds, this results in an excess at $35 \Msun$, as discussed in Section \ref{subsec:excess}. 

    \item The PISN mass gap BHs are a result of super-Eddington accretion in BH+star systems with $5 < q < 10$ (see Section \ref{sec:extended_masses}). These systems are able to merge within the Hubble time due to tidal forces during Roche Lobe Overflow bringing the system together. Radial shrinkage due to mass loss and the efficient accretion by the BH, which flips the mass ratio, allows this interaction to be stable. These high-mass systems create a disconnect between the excess and the cut-off of the primary BH mass distribution, as observed.
    
    \item  Super-Eddington accretion also restricts the amount of mass ratio reversal to 4 per cent with most occurring in systems undergoing CEE (Section \ref{subsec:MRR}). Furthermore, the large amount of mass transferred onto the primary BH could leave an imprint on the spin of the BH, visible during merger. While we are currently unable to predict the spin of the system, we find that most mass is transferred in merging systems with $q \lesssim 0.5$ and high chirp masses, similar to values found by \citet{zevin_2021}, aligning with observations (Section \ref{sec:BHspin}).
    
    \item Besides the standard \textsc{bpass} remnant mass prescription, which injects $10^{51}$ erg into the star to determine the remnant mass, we also implement the \citet{fryer_2012} prescriptions without PPISN. Because the prescriptions are similar in the high CO core mass regime, the $35 \Msun$ and upper mass gap BHs remain, but the low mass regime of the $\MpBH$ distribution and the mass ratio distribution changes significantly with Rapid prescription providing a closer match to the observed distributions.

    \item Completely removing the PPISN and/or PISN prescription only minimally impacts the $\MpBH$ distribution, because the majority of BBH progenitors in \textsc{bpass} do not experience PPISN or PISN.

    \item Quasi-homogeneous evolution is an essential physical process in shaping the $\MpBH$ distribution. Altering the QHE selection does significantly alter the $\MpBH$ distribution, as discussed in Section \ref{sec:QHE_limit}. If high amount of mass are transferred to the companion before the first supernova, further interactions are limited due to QHE. Allowing for less QHE, the rate of the $35 \Msun$ excess increases and shifts it to higher masses. Implementing a detailed determination of the accretion QHE from \citet{ghodla_2022a} based on the amount of material accreted, only slightly alters the final $\MpBH$ distribution.

    \item Because we use detailed stellar models, we model the response of the donor star due to mass loss instead of implementing prescriptions based on the evolutionary phase of the star. We find that the stability of mass transfer depends on metallicity, mass, and age,  similar to \citet{ge_2010, ge_2015, ge_2020, ge_2020a}. For comparison, we show the nature of the envelope for our single star models over a large range of metallicity, ZAMS masses, and ages. Most importantly, the envelope during core-helium burning is often radiative, while rapid population synthesis codes based on \citet{hurley_2002} often assume that these stars have a convective nature \citep[for more detail, see][]{klencki_2021}. As a result, more systems undergo nuclear timescale stable mass transfer in \textsc{bpass}, especially at larger mass ratios between a BH and stellar companion.

\end{enumerate}

While the evolution of single stars is complex and non-linear, evolution of binary stars is even more extreme. All relevant factors have to be taken into account with QHE and the stability and efficiency of mass transfer being essential in understanding the high-mass regime of the primary BH mass distribution. Besides constraining other observable properties of the population synthesis, such as (Ultra-Luminous) X-Ray binaries, future GW observations will be able to further constraint the properties of the merging BBH population and restrict the formation pathways.

\section*{Acknowledgements}
MMB would like to thank Sohan Ghodla, Lieke van Son, Rob Farmer, Stephen Justham, David Hendriks, and Jakub Klencki for the helpful discussions. The authors thank the anonymous reviewer for their insightful comments that have improved this work.
MMB, JJE and HFS acknowledge support by the University of Auckland and funding from the Royal Society Te Apar\={a}ngi of New Zealand Marsden Grant Scheme. We are grateful to the developers of python, matplotlib \citep{hunter_2007}, numpy \citep{harris_2020a} and pandas \citep{reback_2020}.

The \textsc{bpass} project makes use of New Zealand eScience Infrastructure (NeSI) high performance computing facilities. New Zealand's national facilities are funded jointly by NeSI's collaborator institutions and through the Ministry of Business, Innovation \& Employment's Research Infrastructure programme.

\section*{Data Availability}

 The \textsc{bpass} models can be found at \url{https://bpass.auckland.ac.nz/}. The code for plots can be found on our organisational github https://github.com/UoA-Stars-And-Supernovae. The envelope and mass transfer stability data in machine readable format can be found on Github or in the Supplementary Material.



\bibliographystyle{mnras}
\bibliography{references} 




\appendix
\section{Model uncertainties} \label{app:uncertainties}

\textsc{bpass} is based on a limited number of stellar models (250.000) and one concern could be the number of models contributing to the $\MpBH$ distribution. To this end we calculate the number of models after the natal kick (465.047) contributing to each bin in the $\MpBH$ distribution and calculate the Poisson error for each bins. The number of models after the natal kick are larger due to sampling of the natal kick distribution. The resulting Poisson errors are insignificant and, except for the lowest mass bin, are unnoticeable when added to Figure \ref{fig:m1_q_distributions}.

\begin{figure}
    \centering
    \includegraphics[width=\columnwidth]{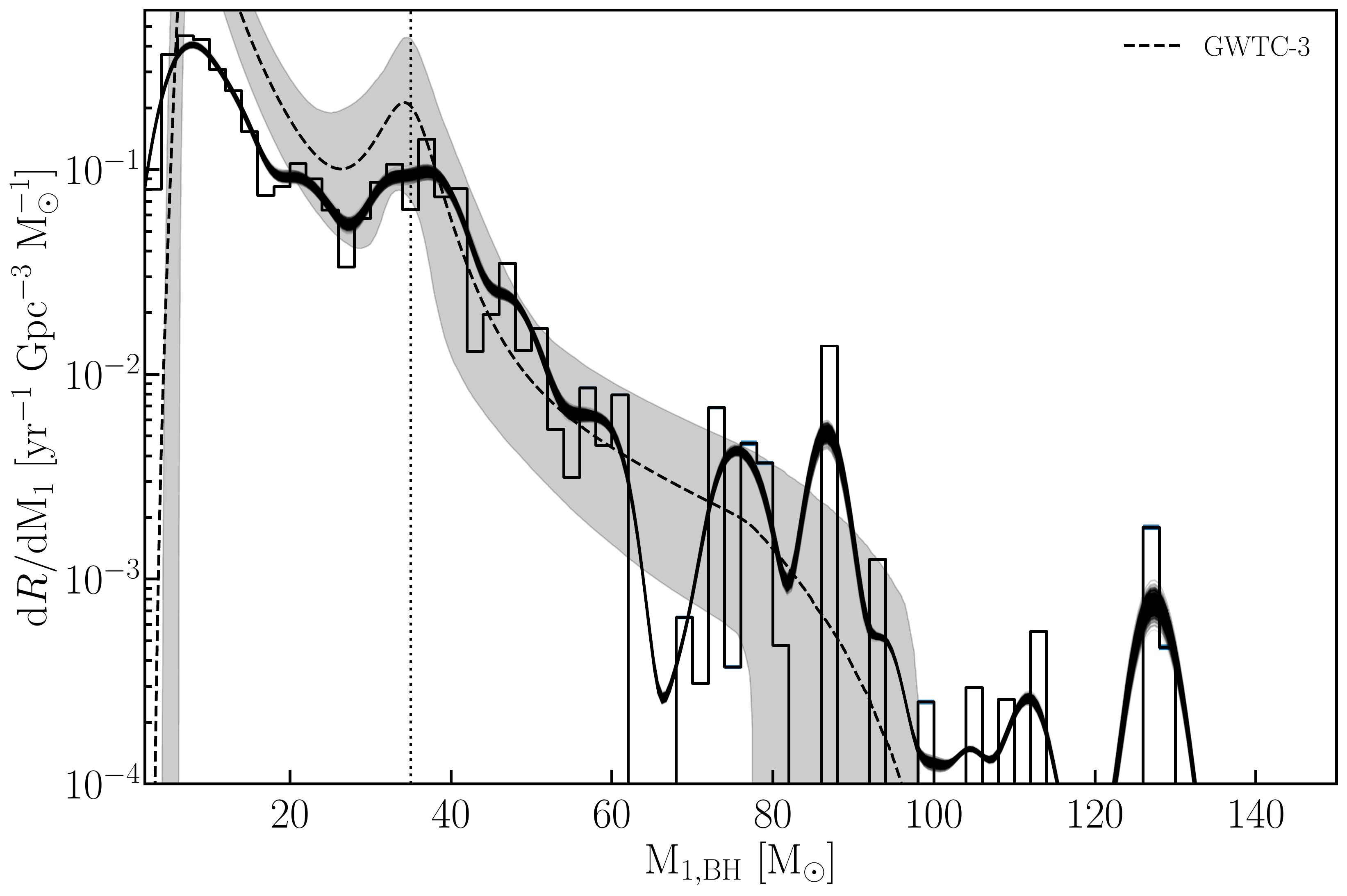}
    \caption{The $\MpBH$ distribution using linear bins with a width of $2 \Msun$. The vertical dotted line indicates $35 \Msun$ and the dashed line with gray area is the GWTC-3 intrinsic population and the 95 per cent confidence interval using their \textsc{pp} model \citep{abbott_2021c}. The thick black line are kernel density estimates from 1000 resampled populations.}
    \label{fig:m1_linear}
\end{figure}

Because the ZAMS mass grid at high masses is sparse in \textsc{bpass}, we have selected a logarithmic binning for the $\MpBH$ space to combine the systems in this regime. In Figure \ref{fig:m1_linear}, we show the $\MpBH$ distribution with linear bins of a $2 \Msun$ width. As can be seen the high-mass regime is sparsely populated, but the number of models contributing to each bin is still results in insignificant errors. Furthermore, the $35 \Msun$ excess and a substructure at $20 \Msun$ are still present.

Instead of binning the \textsc{bpass} output, we can instead apply a kernel density estimate (KDE) to the individual $\MpBH$ merger samples. Using Scott's rule for bandwidth determination, we get the solid black line in Figure \ref{fig:m1_q_distributions}. The KDE fluctuates significantly due to the undersampling of masses in the high-mass regime and small event rates. The $35 \Msun$ peak wider and peaks slightly higher and the substructure at $20 \Msun$ is clearly visible. This indicates that both these features are part of distribution and not a binning artefact. 
Furthermore, we perform 1000 resamplings of the data by redrawing the same number of models as the original $\MpBH$ distribution with each model weighted equally. We run the new populations through the KDE, which show as the thick solid black line in Figure \ref{fig:m1_linear}. Although some statistical uncertainty is present, it does not significantly impact the $\MpBH$ distribution or its features.

The features in the $\MpBH$ distribution remain when changing bins and when applying a kernel density estimation. While the high $\MpBH$ range is undersampled, sufficient models contribute to each mass for the statistical error to be negligible. Of course, this does not take into account uncertainty based on the implemented physics.

\section{Homology} \label{app:homology}
In the context of mass transfer the thermal timescale of the star are important because it limits the accretion onto the companion star. Within \textsc{bpass} this limit is $\dot{\mathrm{M}} = \mathrm{M}_2 / \tau_\mathrm{KH}$.
Thus, the timescale at which the primary loses the mass and at which the secondary star can accrete this material determines the mass lost from the system. To be able to get massive BHs, it is essential that material remains in the system and that the timescales are similar. To understand the behaviour of the stellar models, we use homology relations to determine the dependence of the thermal timescale on mass between different stars \citep[Found in most stellar structure text books, such as][]{eldridge_2019a}. The thermal timescale is proportional to the mass, luminosity and radius of the star, as per Equation \ref{eq:KH}.
\begin{equation} \label{eq:KH}
    \tau_\text{KH} \propto \frac{M^2}{RL}
\end{equation}

To determine the dependence solely on mass, we setup a few basic scaling relations that approximate the behaviour of massive stars before turning our attention to specific regimes.
From the first two stellar structure equations, we get the following main homology relations in which $\rho_c$ and $P_c$ are the central temperatures of the star with mass $M$ and radius $R$.
\begin{equation} \label{eq:h1}
    \rho_c \propto \frac{M}{R^3}
\end{equation}

\begin{equation} \label{eq:h2}
    P_c \propto \frac{M^2}{R^4}
\end{equation}

We use the stelar structure equation of radiative energy transport and electron scattering ($\kappa \propto 1+X$) for the star. After applying Equation \ref{eq:h1}, we get Equation \ref{eq:h3}.
\begin{equation} \label{eq:h3}
    \frac{T_c}{R} \propto \frac{\kappa \rho_c L}{R^2 T_c^3} \longrightarrow L \propto T_c^4\frac{R^4}{M}\frac{1}{1+X},
\end{equation} 
where $T_c$ is the central temperature, $L$ the luminosity of the star, and $X$ the hydrogen fraction.

Finally, we use the energy generation of the star to get Equation \ref{eq:h4}. Since the main energy generation for massive stars is through the CNO-cycle, we have $\nu \approx 18$.

\begin{equation} \label{eq:h4}
    \frac{L}{R} \propto R^2\rho_c^2 T_c^{\nu} \longrightarrow L \propto \frac{M^2}{R^3} T_c^{18}
\end{equation}

\subsection{Gas pressure dominated}
At the low end of massive stars, the star is dominated by gas pressure and the equation of state follows the ideal gas law, where $\mu$ is the mean molecular weight in the star.
\begin{equation} \label{eq:gas}
    P_c \propto \frac{\rho_c T_c}{\mu} \longrightarrow T_c \propto \mu \frac{M}{R}
\end{equation}
This allows us to transform Equations \ref{eq:h3} into
\begin{equation} \label{eq:h5}
    L\propto \frac{\mu^4}{1+X} M^3
\end{equation}

And combine it with \ref{eq:h4} to get:
\begin{equation} \label{eq:h6}
    L \propto \mu^{18} \frac{M^{20}}{R^{21}} \longrightarrow \frac{\mu^4}{1+X} M^3 \propto \mu^{18}  \frac{M^{20}}{R^{21}} 
\end{equation}
Eventually, we get the relation between radius and mass. 
\begin{equation} \label{eq:h7}
    R \propto \mu^{18/21} (1+X) M^{17/21}
\end{equation}

Combining Equations \ref{eq:h5}, \ref{eq:h6}, \ref{eq:h7} with Equation \ref{eq:KH}, we find that for the ideal gas equation of state, the thermal timescale scales as follows:
\begin{equation}
    \tau_\text{KH} \propto \frac{M^2}{M^{17/21}M^3} \propto M^{-38/21} \propto M^{-1.81}
\end{equation}

\subsection{Radiation pressure dominated}
At higher masses, radiation pressure plays an essential role in supporting the star. Thus, the ideal gas equation of state is no longer valid. While the actual dynamics will be a combination of radiation and gas pressure, we can look at the extreme example with only radiation pressure. Instead of Equation \ref{eq:gas}, we get the following relation for the central temperature
\begin{equation}
    P_c \propto T^4_c \longrightarrow T^4_c \propto \frac{M^2}{R^4}
\end{equation}
This gives us the following equations for the luminosity and radius.

\begin{equation}
    L \propto \frac{1}{1+X} M
\end{equation}

\begin{equation}
    L \propto \frac{M^{10}}{R^{21}} \longrightarrow \frac{M}{1+X} \propto \frac{M^{10}}{R^{21}} \longrightarrow R \propto M^{10/21}
\end{equation}
Using them we find the relation between the thermal timescale and mass in radiation pressure only regime.
\begin{equation}
    \tau_\text{KH} \propto M^{11/21} \propto M^{0.52}
\end{equation}

\subsection{Relations within the \textsc{bpass} models}

The previous two sections cover two extreme cases. In reality, both the gas and radiation pressure contribute to the total internal pressure in the star. However, the difference in thermal timescale dependence on the mass is significant and means that at higher masses, where radiation pressure dominates, the thermal timescale becomes constant with mass.

\begin{figure}
    \centering
    \includegraphics[width=\columnwidth]{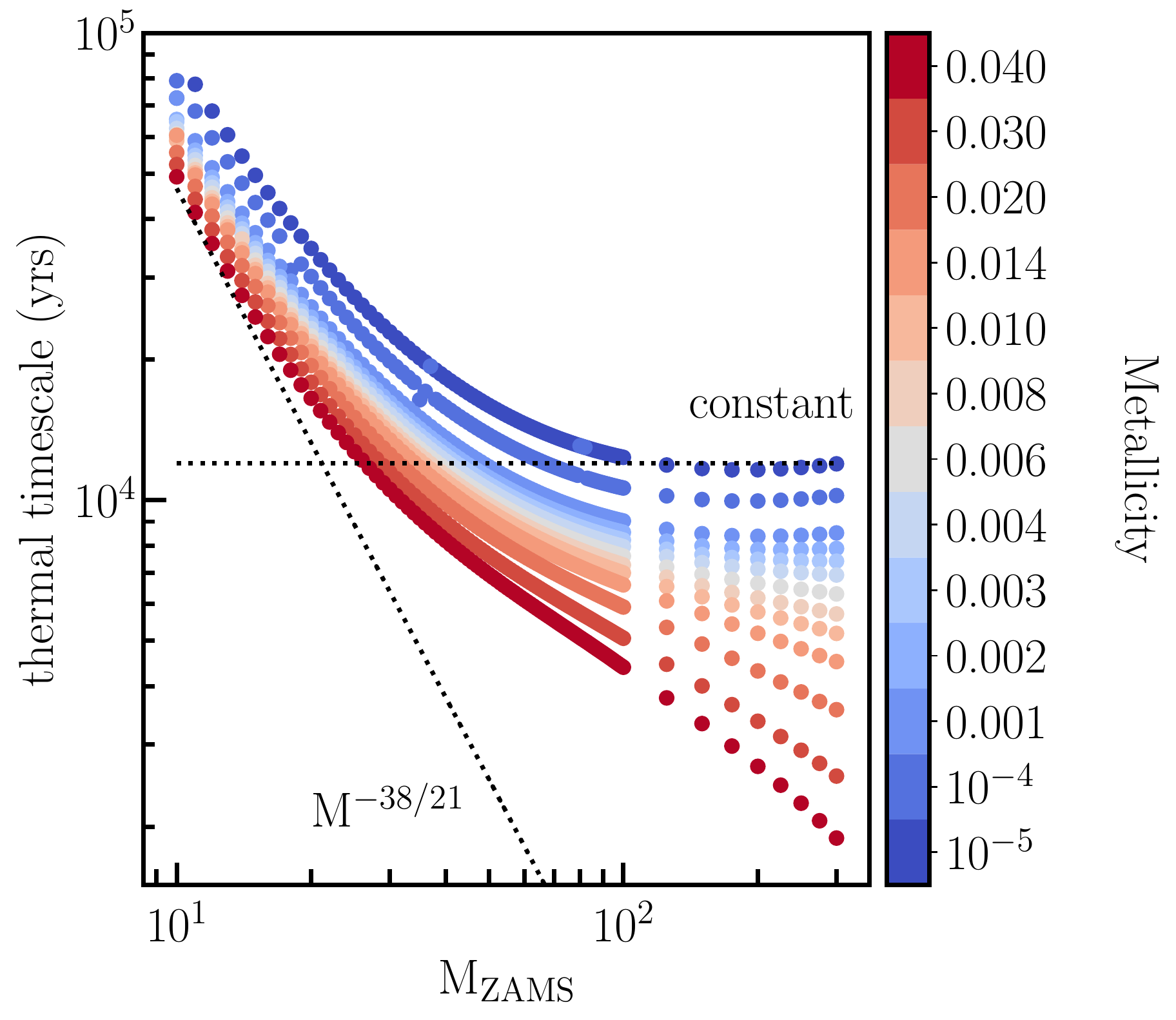}
    \caption{The thermal timescale at ZAMS of all \textsc{bpass} single star model per metallicity. The dashed lines indicate a $\tau_\text{KH} \propto M^{-38/21}$ relation and the thermal timescale being constant.}
    \label{fig:timescale}
\end{figure}

To verify this, we plot the thermal timescale at ZAMS of all \textsc{bpass} single star models in Figure \ref{fig:timescale}. At sub-solar metallicity, where BBH mergers are though to originate from \citep{belczynski_2010, stevenson_2017, stevenson_2019, santoliquido_2021}, the thermal timescale becomes close to constant for high masses. Thus, if mass is lost on a thermal timescale, most of it can be accreted by the companion, because the timescales are similar. 

At higher metallicities, the timescales no longer become constant, which is most likely due to the change in the opacity of the star.

\bsp	
\label{lastpage}
\end{document}


\label{firstpage}
\pagerange{\pageref{firstpage}--\pageref{lastpage}}
\maketitle



\appendix
\section{Nature of the envelope from single stars}

During mass transfer the radiative or convective nature of the envelope of the donor star can influence how the interaction progresses. Generally, a convective envelope star expands, while a star with a radiative envelope shrinks as a response to mass loss.

Because of the large grid of detailed stellar models required for population synthesis, we can determine the nature of the envelope of single stars over a large range of initial masses and metallicities using our adapted \textsc{bpass} version of the Cambridge \textsc{stars} code. This can be useful in determining the stability of mass transfer in binary interactions and comparison against other other work.

The \textsc{bpass stars} code only provides the mass coordinates of convective boundaries, but not which regions are convective or radiative. We determine this using the semi-convective boundaries, which are included in the mass coordinates. We do have to assume that the core of the star is convective at the start of the model. In the mass regime considered here ($\text{M}_\text{ZAMS} > 10 \Msun$), this is reasonable, but does not necessarily hold at smaller masses \citep{eldridge_2019a}.

We define a convective envelope if a star has a convective region that is at most $0.1 \Msun$ below the surface of the star and is at least $1 \Msun$ deep. Our automatic tagging is not guaranteed to work in all situations due changes in the convective boundaries array that are numerical and not physical. Thus, we check every model to determine if the tagging of the convective envelope was successful. While most models are tagged successfully, if the tagging fails, we adjust the convective regions manually.

\begin{figure*}
    \centering
    \includegraphics[width=\textwidth]{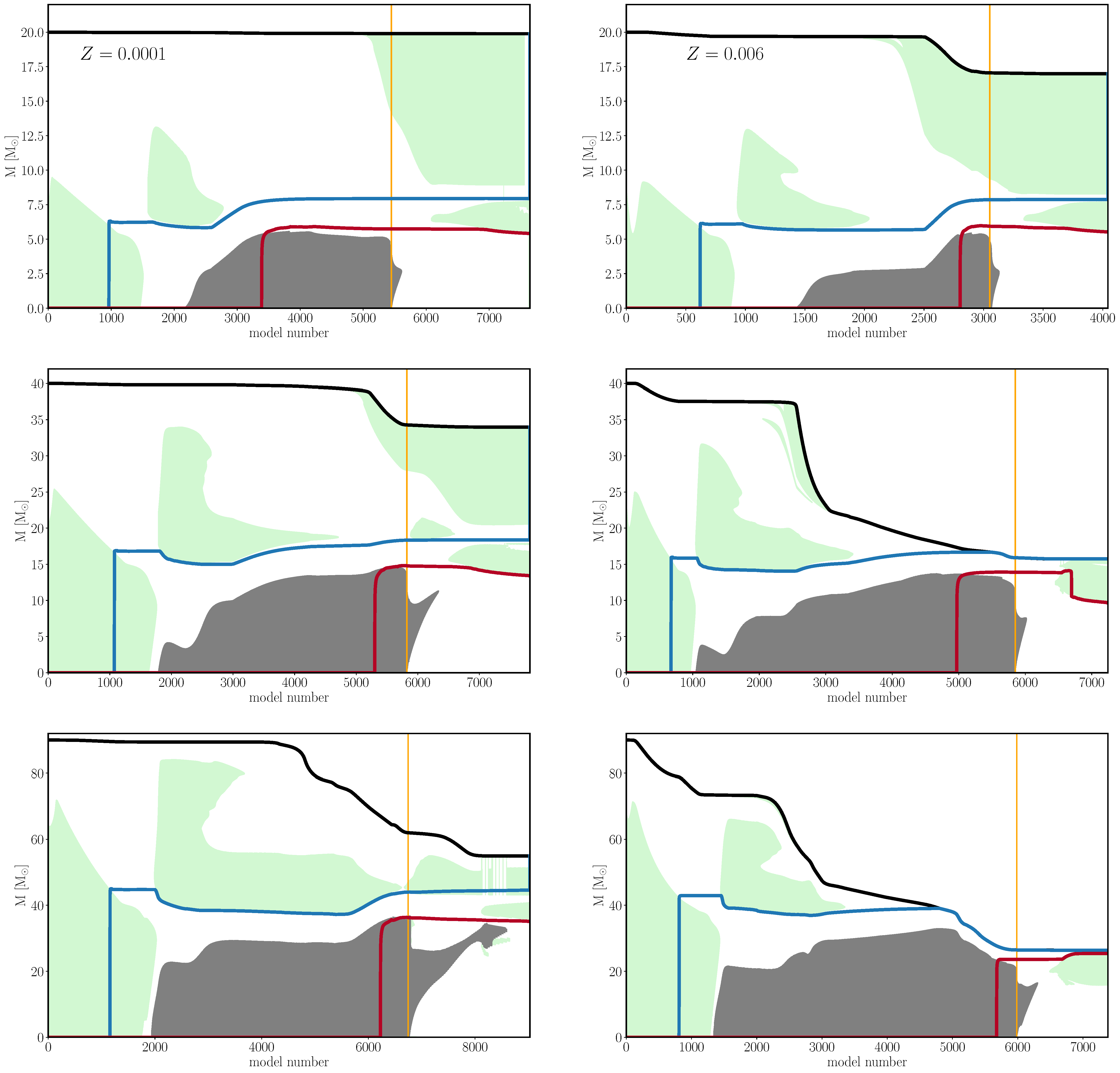}
    \caption{The Kippenhahn plots of a 20, 40, and 90 $\Msun$ star at $Z=0.0001$ and $Z=0.006$. The total mass, Helium core mass, and Carbon-Oxygen core mass are indicated in black, blue, and red, respectively. The convective regions are green shaded. The grey are is the envelope convective zone tagged by the automated system. We indicate the end of core helium burning by the orange vertical line. The automatic tagging can be seen to fail in later stages of the $20 \Msun$ and $90 \Msun$ model (past the orange line).}
    \label{fig:kippenhahn_examples}
\end{figure*}

Figure \ref{fig:kippenhahn_examples} shows six Kippenhahn plots for stars at different masses and metallicities. The grey area is the automatically tagged area, which failed in the $20 \Msun$ and $90 \Msun$ at $Z=0.0001$ at the end of the model (past the orange line). Because the tagging does not work well in later stages of the evolution, we only look at the envelope structure until the end of core Helium burning. We define the end of core Helium burning, when the location of the convective core during core-Helium burning moves above $0.1 \Msun$, which results in the orange line in Figure \ref{fig:kippenhahn_examples}. For models that did not reach this, we use the age of the closest model as the end of core Helium burning.

After manually altering the incorrectly tagged regions, we explore how the nature of the envelope changes over initial mass and metallicity.

\begin{figure}
    \centering
    \includegraphics[width=\columnwidth]{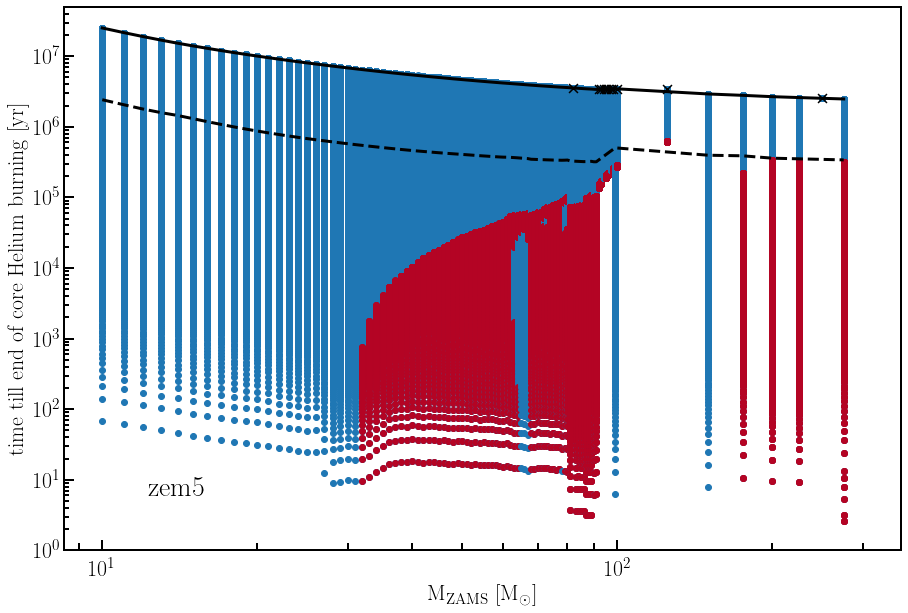}
    \caption{The radiative (blue) or convective (red) nature of the envelope from a single star over different ZAMS masses up to the end of core Helium burning. The end of the Main-Sequence is indicated (black dashed) and so are models that did not converge (black crosses).}
    \label{fig:convective_nature_zem5}
\end{figure}

Figure \ref{fig:convective_nature_zem5} shows the nature of the envelope over initial masses at $Z=0.00001$ till the end of core helium burning. Red areas indicate a convective envelope, while blue indicates a radiative envelope. Models that do not reach the end of core helium burning are indicated with a black cross and the final time of the previous (lower mass) star is used to approximate its end point. The end of the ZAMS is indicated by the dashed line and is found by looking at when the helium luminosity starts to increase. Because we are limited by the indicator available in \textsc{stars}, the end of core Hydrogen and Helium burning might not be the actual moment this occurs, but should be similar.

On the Main-Sequence all stars in this mass regime are radiative. Post Main-Sequence the nature of the envelope depend on the initial mass and metallicity of the star. At $Z=0.0001$ below $\text{M}_\text{ZAMS} \leq 21 \Msun$, the star remains radiative. Above this limit, the star becomes convective during the core helium burning phase and more massive stars become convective earlier in the phase.

As metallicity increases (see additional Figures), the low mass stars do become convective after entering the core-helium burning phase. At the same time, higher mass stars become radiative during core helium burning after a short phase of convection after the end of the MS. Eventually, stars below $26 \Msun$ are mostly convective during core-helium burning, while higher mass stars are radiative, agreeing with results found by \citet{klencki_2021}. Indicating that interactions during this phase are more stable than estimated by rapid population synthesis codes. With a shorter period of convection as the ZAMS mass increases until it eventually disappears in the high mass regime. This is most likely a result of strong winds stripping most of the star and no convective envelope forming, as can be seen the Kippenhahn diagram of the $90 \Msun$ star at $Z=0.006$ in Figure \ref{fig:kippenhahn_examples} (bottom right).

\begin{figure}
    \centering
    \includegraphics[width=\columnwidth]{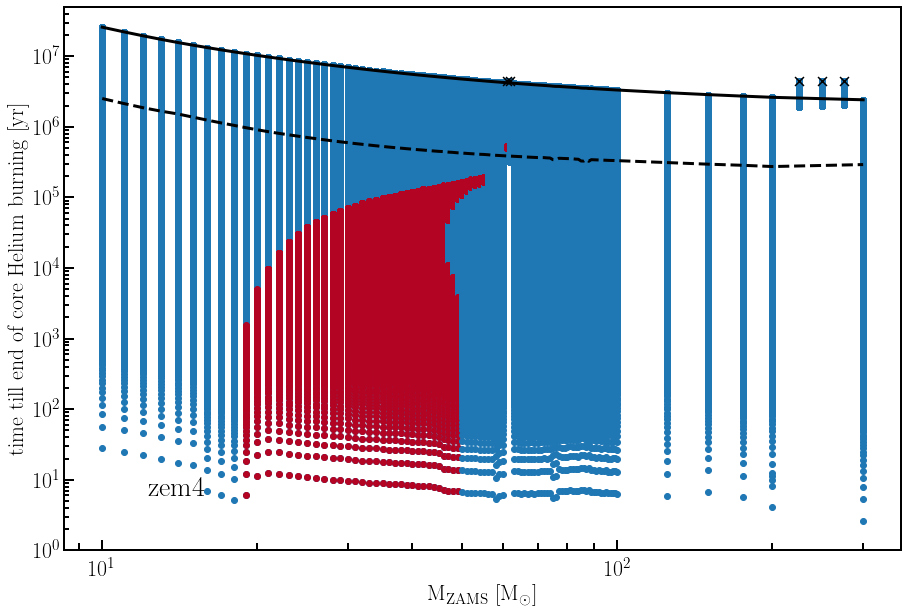}
    \caption{Same as Figure \ref{fig:convective_nature_zem5}, but for $Z=0.0001$.}
\end{figure}
\begin{figure}
    \centering
    \includegraphics[width=\columnwidth]{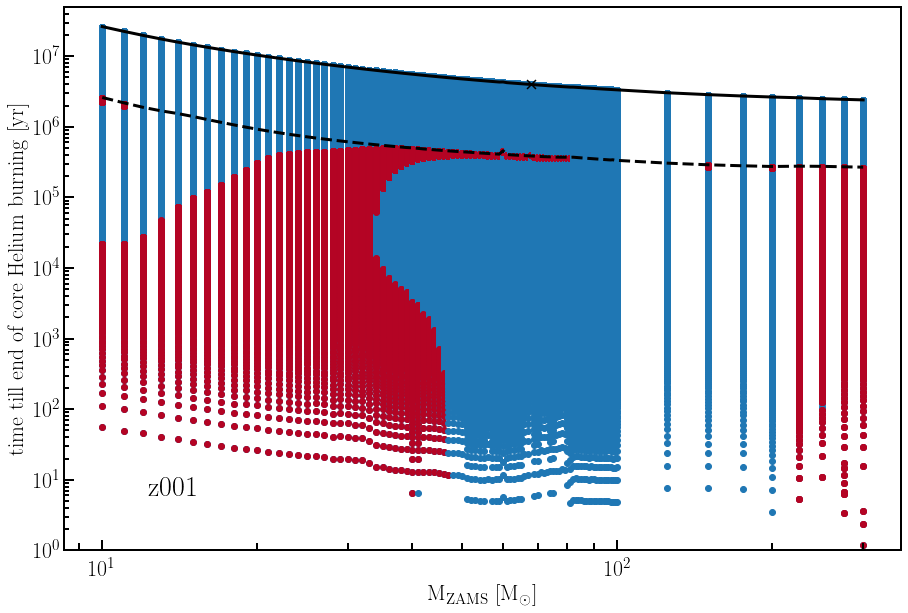}
    \caption{Same as Figure \ref{fig:convective_nature_zem5}, but for $Z=0.001$.}
\end{figure}
\begin{figure}
    \centering
    \includegraphics[width=\columnwidth]{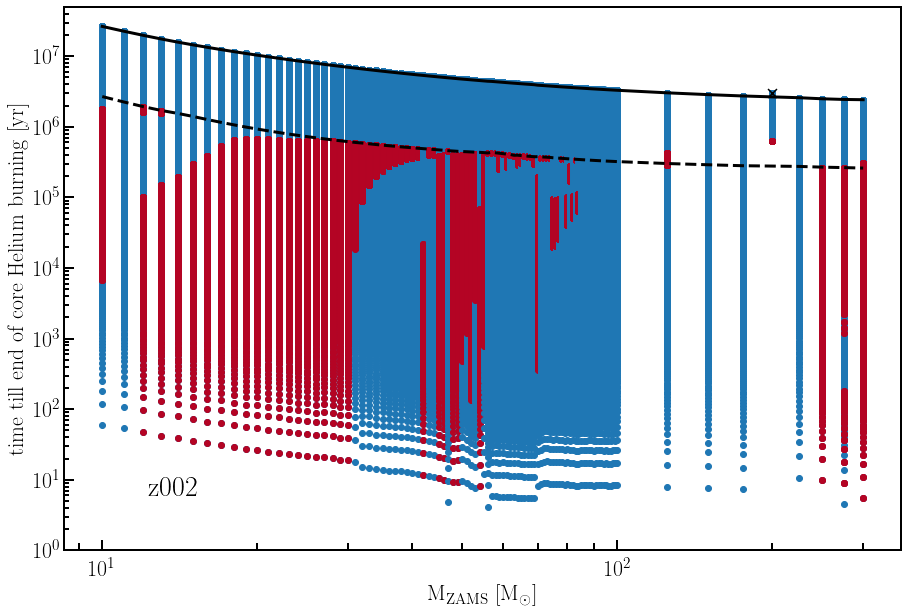}
    \caption{Same as Figure \ref{fig:convective_nature_zem5}, but for $Z=0.002$.}
\end{figure}
\begin{figure}
    \centering
    \includegraphics[width=\columnwidth]{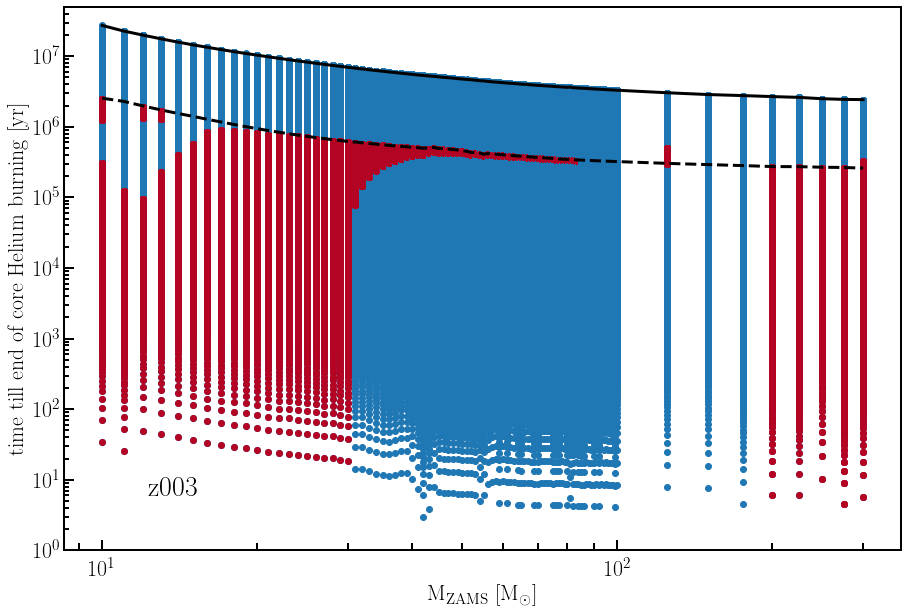}
    \caption{Same as Figure \ref{fig:convective_nature_zem5}, but for $Z=0.003$.}
\end{figure}
\begin{figure}
    \centering
    \includegraphics[width=\columnwidth]{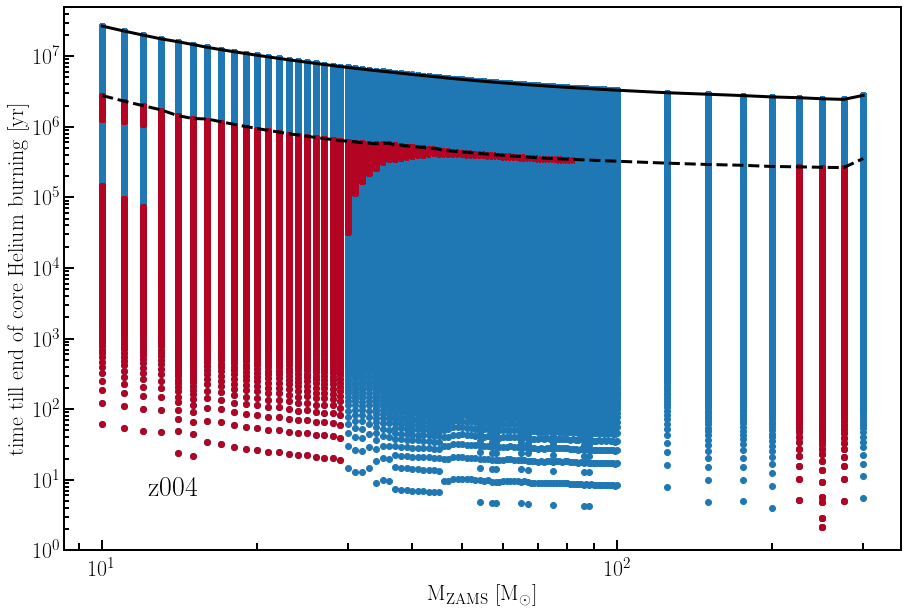}
    \caption{Same as Figure \ref{fig:convective_nature_zem5}, but for $Z=0.004$.}
\end{figure}
\begin{figure}
    \centering
    \includegraphics[width=\columnwidth]{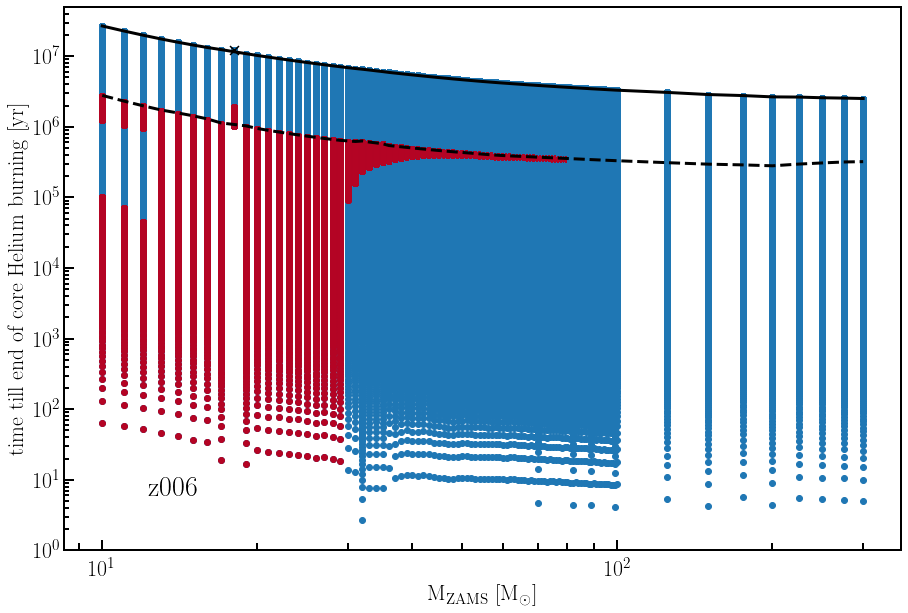}
    \caption{Same as Figure \ref{fig:convective_nature_zem5}, but for $Z=0.006$.}
\end{figure}
\begin{figure}
    \centering
    \includegraphics[width=\columnwidth]{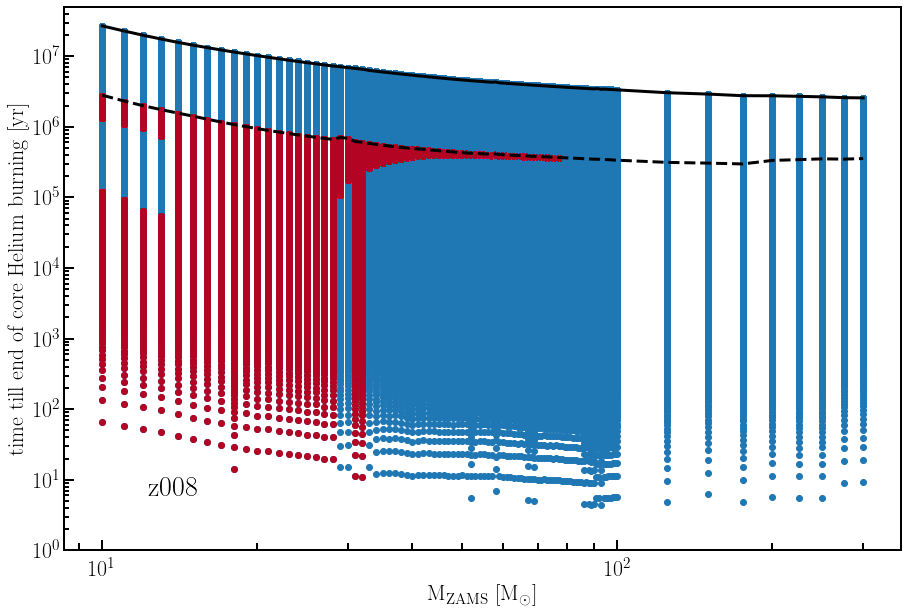}
    \caption{Same as Figure \ref{fig:convective_nature_zem5}, but for $Z=0.008$.}
\end{figure}
\begin{figure}
    \centering
    \includegraphics[width=\columnwidth]{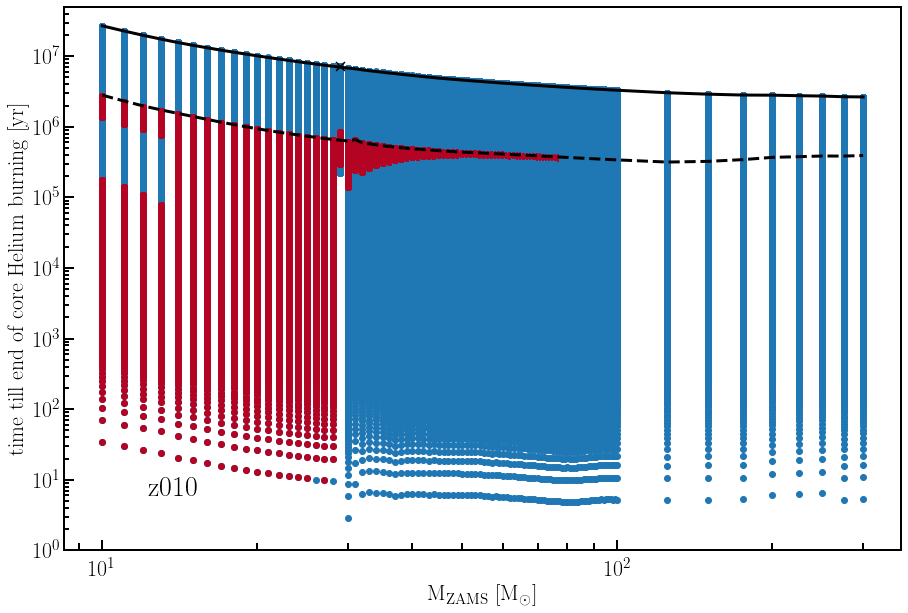}
    \caption{Same as Figure \ref{fig:convective_nature_zem5}, but for $Z=0.01$.}
\end{figure}
\begin{figure}
    \centering
    \includegraphics[width=\columnwidth]{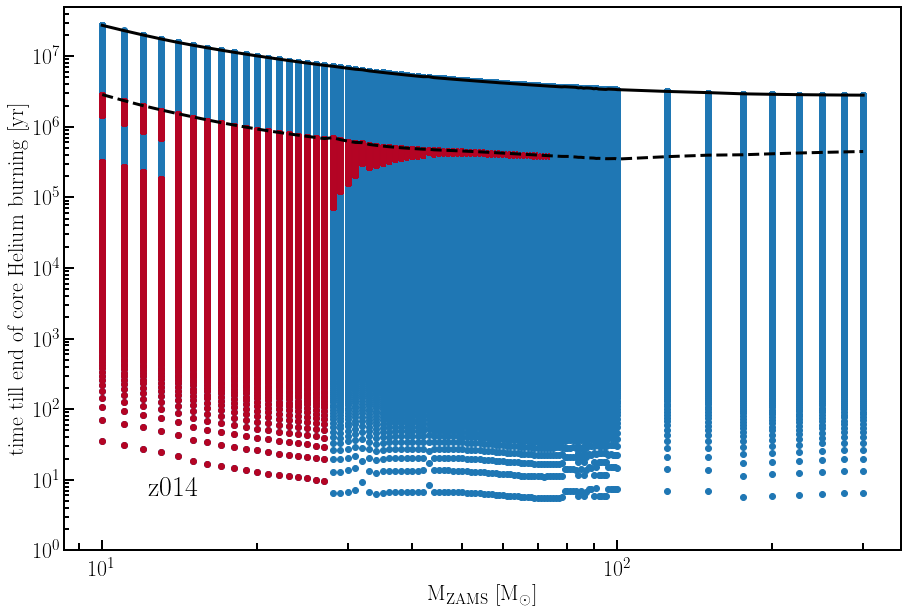}
    \caption{Same as Figure \ref{fig:convective_nature_zem5}, but for $Z=0.014$.}
\end{figure}
\begin{figure}
    \centering
    \includegraphics[width=\columnwidth]{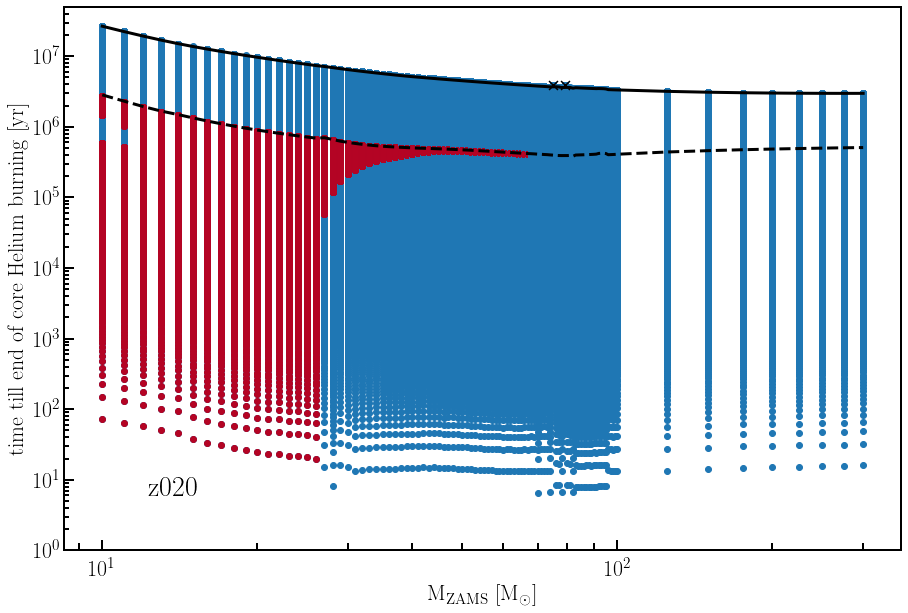}
    \caption{Same as Figure \ref{fig:convective_nature_zem5}.}
\end{figure}
\begin{figure}
    \centering
    \includegraphics[width=\columnwidth]{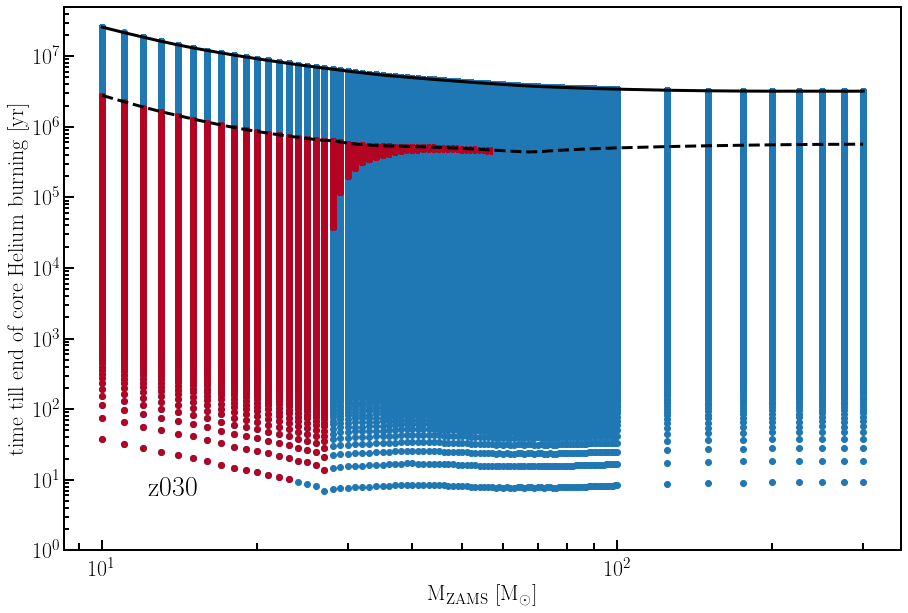}
    \caption{Same as Figure \ref{fig:convective_nature_zem5}, but for $Z=0.03$.}
\end{figure}

\begin{figure}
    \centering
    \includegraphics[width=\columnwidth]{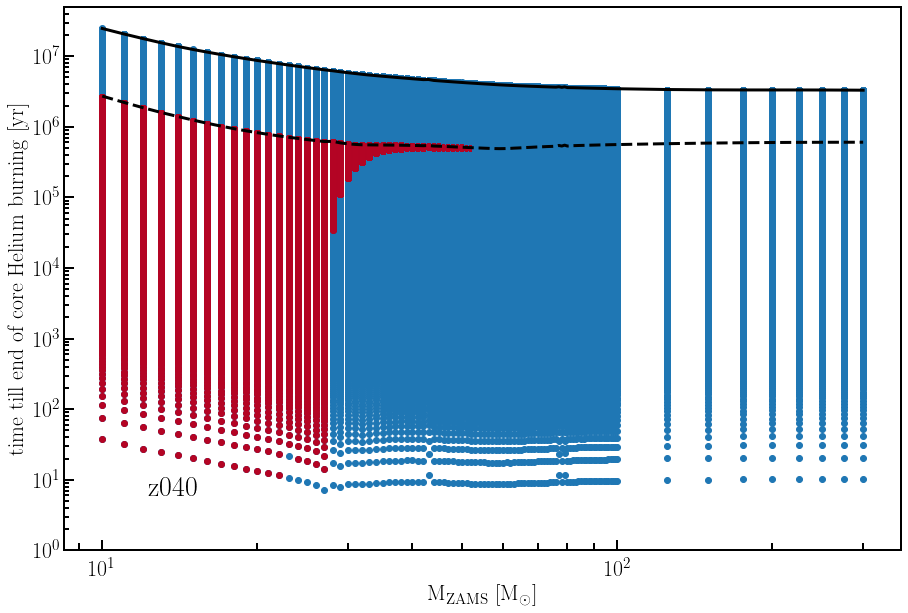}
    \caption{Same as Figure \ref{fig:convective_nature_zem5}, but for $Z=0.04$.}
\end{figure}

\section{Stability of Mass Transfer in \textsc{bpass}}

The stability of mass transfer can significantly alter the outcome of the binary evolution, and the resulting population properties, as shown in Section 4.1. Because the stellar evolution of \textsc{bpass} is based on the Cambridge \textsc{stars} code, the response of the donor star to mass loss from Roche Lobe Overflow can be determined in the code. As a result, no critical mass ratio or $\zeta$ parameters are required in determining the stability of the mass transfer. Instead, we impose a basic physical definition of unstable mass transfer. When the radius of the donor star becomes larger than the separation of the system, the star star undergoes a Common Envelope (CE) phase. This prescription might be too lenient, but allows us to take into account the metallicity, mass, stellar structure, and binary parameters of the systems, which can all effect the stability of the mass transfer \citet{ge_2010, ge_2015, ge_2020, ge_2020a}. Moreover, it allows for mass transfer before and after the CE phase, which can affect the outcome of the interaction. 

While the actual binary interactions are complex and can contain multiple phases of Roche Lobe overflow, each which can be stable or unstable, we identify a binary system as CEE if any of these phases are unstable. As a result, in stable mass transfer (SMT) systems all interactions were stable. We do not indicate merger systems, however, by design these only occur in the CEE channel.

\textsc{bpass} contains two types of binary models: the primary models, where the primary (more massive) star is evolved in detail up to its supernova, and the secondary models, where the companion is evolved in detail with a compact object, if the system remained bound. We describe this process and the stellar evolution in more detail in Section 2. Because the accretion onto a star is different than onto a compact object, we look at each model type separately. The data and a script to create additional stability grids plots for the primary or secondary models is provided.

\subsection{Primary models}

In the primary models, material is transferred from the primary star onto the companion, whose accretion is limited by its thermal timescale. Additional material is lost along with its angular momentum and is treated as lost through stellar winds.
\begin{figure*}
    \centering
    \includegraphics[width=\textwidth]{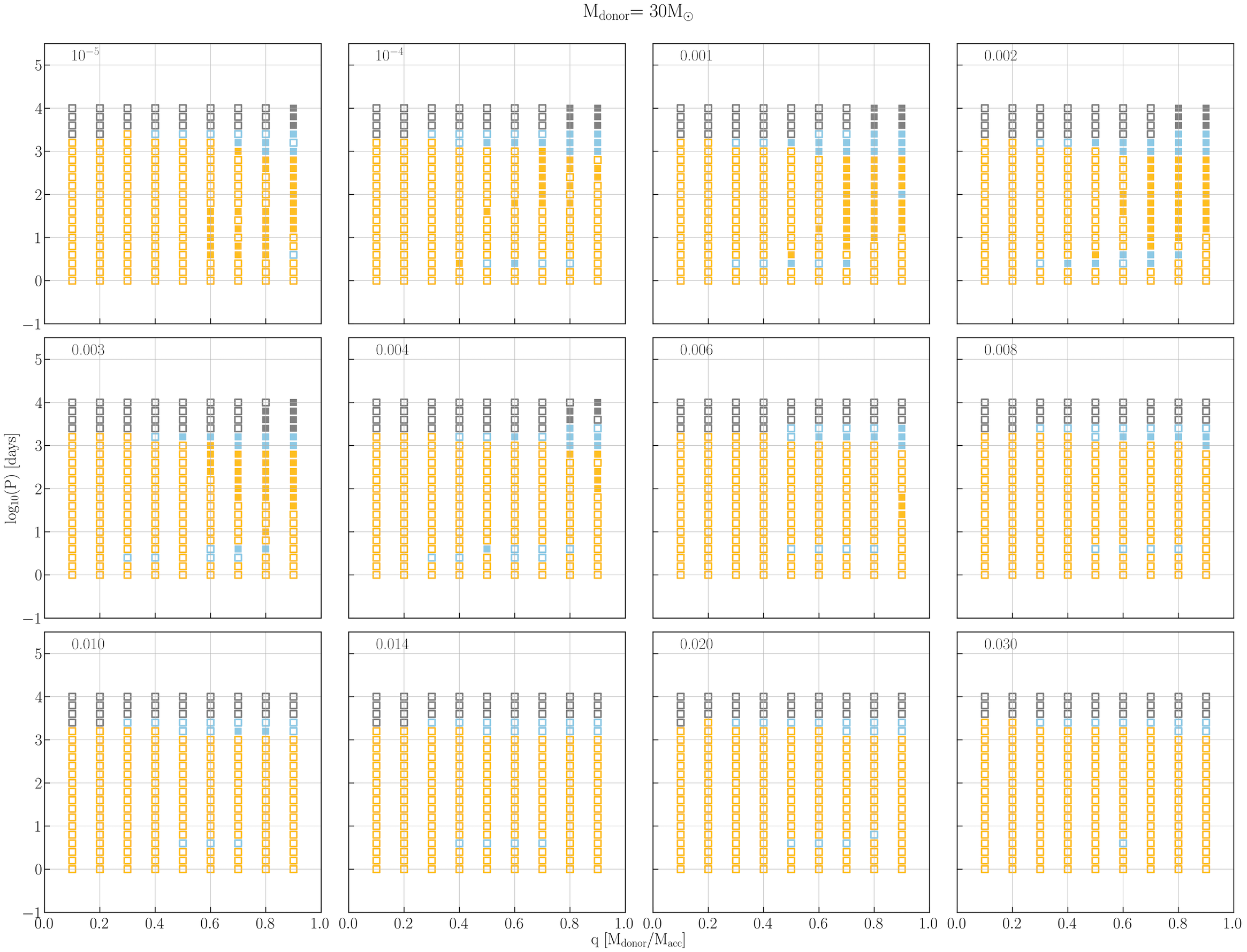}
    \caption{The stability of mass transfer per model in \textsc{bpass} for a $30 \Msun$ donor star with different companion masses for 12 of the 13 metallicities available in \textsc{bpass}. The filled squares indicate models that contribute to the BBH merger rate at $z=0$, while the hollow squares do not. Grey indicates that the model does not interact, yellow that the system undergoes a CE, and blue that the system only experiences SMT.}
    \label{fig:primary1}
\end{figure*}
Since the donor mass, mass ratio and orbital parameters all determine the outcome of the interaction, we will introduce their effect one by one. Figure \ref{fig:primary1} shows the stability of a $30 \Msun$ donor star over the initial mass ratios available in \textsc{bpass} for 12 of the 13 BPASS metallicities over the period of the system. The period is often related to the evolutionary phase of the system, e.g. short period systems interaction on the main-sequence, while large period systems are either interact when expanding during later stages of its evolution. If the period is too large, no interactions will take place.
If no interaction takes place, we indicate this in gray. SMT and CEE are indicated in blue and yellow, respectively. If a model has contributed to the BBH merger rate at $z=0$, the square is filled.

Starting at $Z=0.00001$ in \ref{fig:primary1} at $q=0.9$ most interactions undergo a CEE phase, only at periods around $\sim 2.8$ do these interact stably. Further increasing the period, leads to no interactions. The increased stability in later stages of the evolution are similar to results found by \citet[][(2015, 2020, 2020a)]{ge_2010}. Most likely a result of lower density end temperature of the envelope during this phase of the evolution \citep{ge_2020}.
\begin{figure*}
    \centering
    \includegraphics[width=\textwidth]{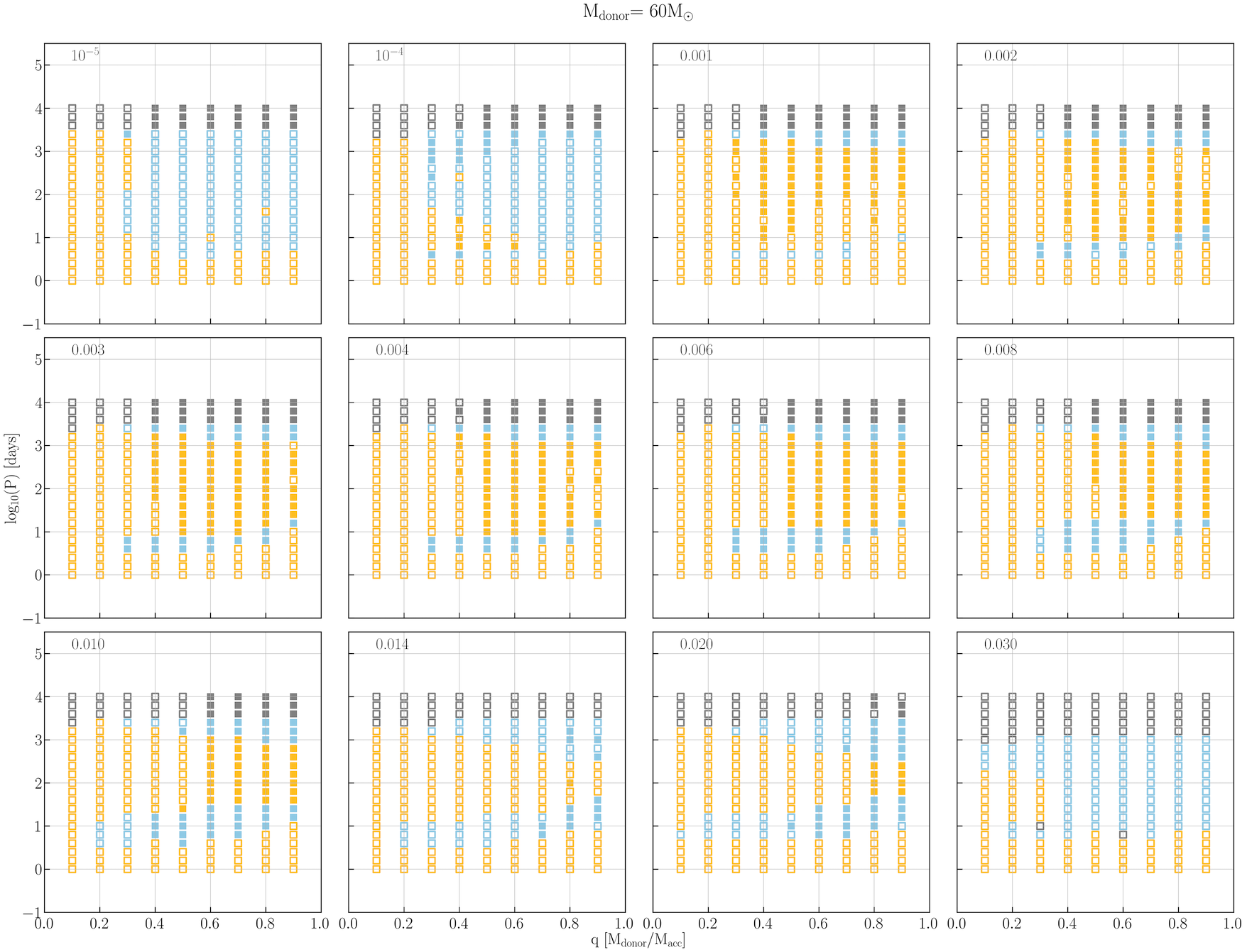}
    \caption{The stability of mass transfer per model in \textsc{bpass} for a $60 \Msun$ donor star with different companion masses for 12 of the 13 metallicities available in \textsc{bpass}. The filled squares indicate models that contribute to the BBH merger rate at $z=0$, while the hollow squares do not. Grey indicates that the model does not interact, yellow that the system undergoes a CE, and blue that the system only experiences SMT.}
    \label{fig:primary2}
\end{figure*}

As the mass ratio decreases, the difference between primary and secondary masses increases and the system is more likely to undergo unstable mass transfer, because mass transfer and tidal synchronisation brings the system close together.

Increasing the metallicity of the system, allows for more stable mass transfer to occur during later phases of the evolution. This is a result of additional mass loss through stellar winds shrinking the orbit. This effect gets less as the mass ratio decreases. Furthermore, there is a slight increase in the stability of short period binaries, as a result of the metallicity increase.

Finally, we can alter the donor mass. Figure \ref{fig:primary2} shows the stability for a $60 \Msun$ donor star. As we increase it, more SMT is allowed on the main-sequence and the post main-sequence. Interactions on the Hertzsprung are often unstable, most likely due to the star evolving on a thermal timescale. Furthermore, larger difference between the donor and companion mass are stable. These observations are similar to the thermal critical mass ratio evolution found by \citet{ge_2020}.

\subsection{Secondary models}

In the secondary binary models, accretion occurs onto a compact object companion. For NS we limit the accretion by the Eddington Luminosity, while we allow for super-Eddington accretion onto BHs.
\begin{figure*}
    \centering
    \includegraphics[width=\textwidth]{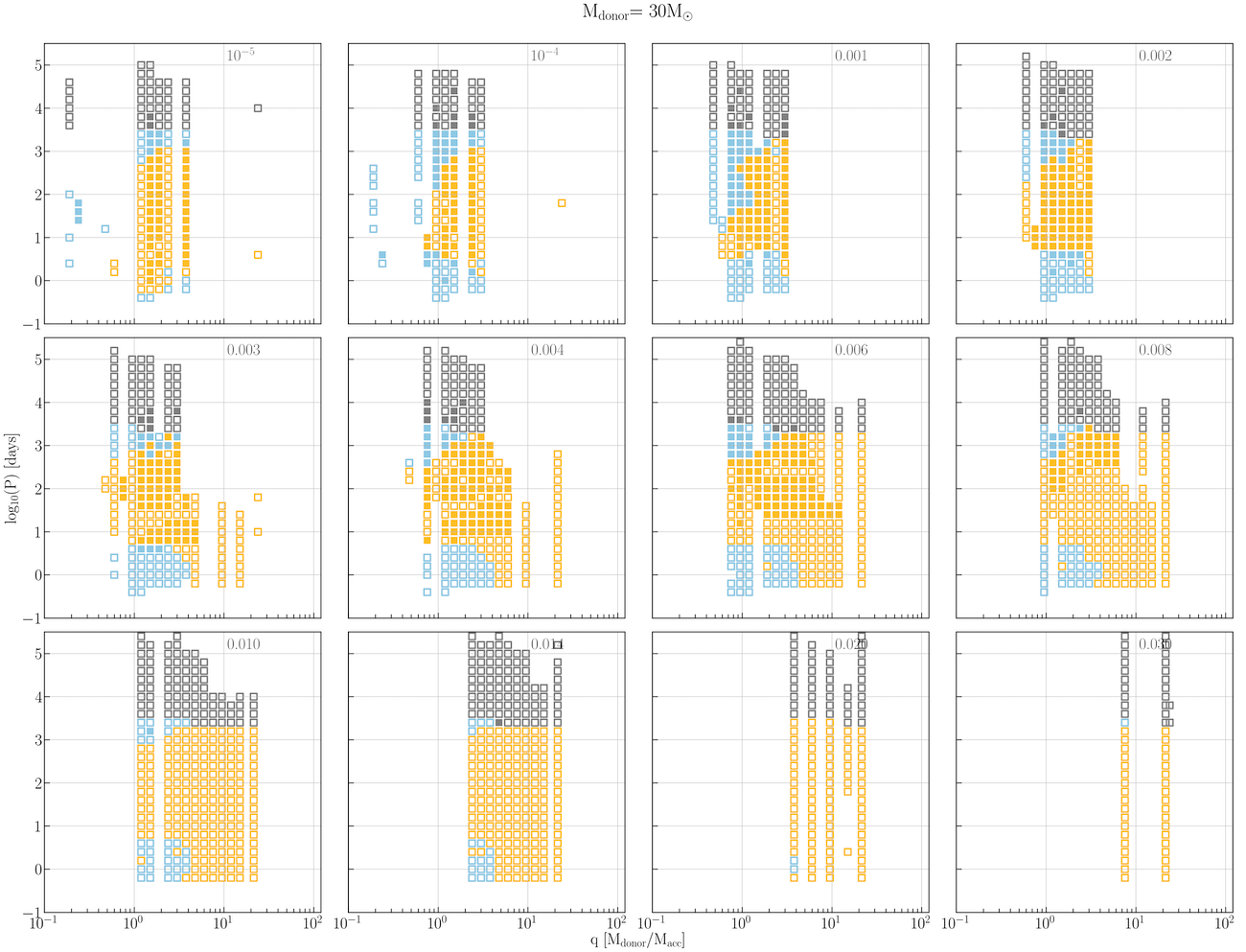}
    \caption{The stability of mass transfer per model in \textsc{bpass} for a $30 \Msun$ donor star with different compact object companion masses for 12 of the 13 metallicities available in \textsc{bpass}. The filled squares indicate models that contribute to the BBH merger rate at $z=0$, while the hollow squares do not. Grey indicates that the model does not interact, yellow that the system undergoes a CE, and blue that the system only experiences SMT.}
    \label{fig:secondary1}
\end{figure*}

We see similar trends in the stability as with the primary models. Figure \ref{fig:secondary1} show a $30 \Msun$ donor star and the possible companions available in \textsc{bpass}. At low metallicity stable interactions are possible on the Main-Sequence, in wide period system (or late stage evolution), and if the compact object is more massive than the donor star ($q < 1$). Initially, with increasing metallicity more SMT is possible. However, if we continue to increase the it, more and more systems undergo CEE and SMT is very restricted.

\begin{figure*}
    \centering
    \includegraphics[width=\textwidth]{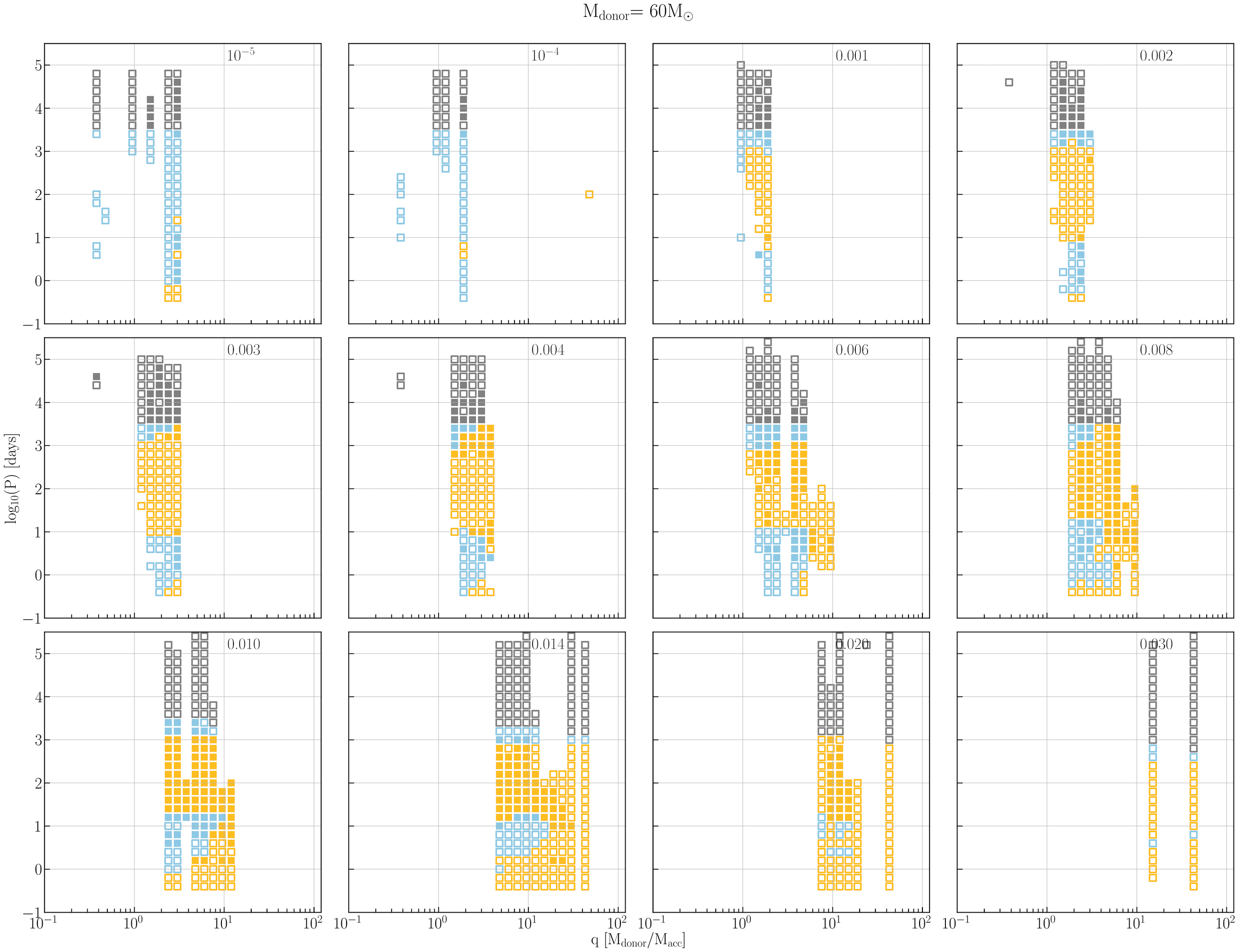}
    \caption{The stability of mass transfer per model in \textsc{bpass} for a $60 \Msun$ donor star with different compact object companion masses for 12 of the 13 metallicities available in \textsc{bpass}. The filled squares indicate models that contribute to the BBH merger rate at $z=0$, while the hollow squares do not. Grey indicates that the model does not interact, yellow that the system undergoes a CE, and blue that the system only experiences SMT.}
    \label{fig:secondary2}
\end{figure*}

Increasing the donor mass to $60 \Msun$ in Figure \ref{fig:secondary2}, once again increases the stability of mass transfer. Albeit sparsely populated, the low metallicity systems mostly undergo SMT. While at higher metallicities, the MS and late stage evolution interaction are now stable up to quite high mass ratios. At the same time, stellar winds have increased, which limits the interactions and reduced the maximum period at which systems interact.

\bibliographystyle{mnras}
\bibliography{references} 

\bsp	
\label{lastpage}